\documentclass[11pt,a4paper]{article}
\pdfoutput=1
\usepackage{jcappub}

\usepackage[utf8]{inputenc}
\usepackage[T1]{fontenc}
\usepackage{graphicx,amssymb,amsmath,wasysym}
\usepackage{xcolor}
\usepackage{slashed}

\newcommand{\mx}{m}

\newcommand{\tk}{T_\text{k}}
\newcommand{\tpk}{{T'_\text{k}}}

\newcommand{\gs}{g_\star}

\newcommand{\teq}{T_\text{eq}}
\newcommand{\zp}{{V}}
\newcommand{\xo}{{\chi_1}}
\newcommand{\xt}{{\chi_2}}
\newcommand{\gd}{g_V}
\newcommand{\mzp}{m_\zp}

\newcommand{\NN}[6]{\langle #1#2#3#4\to #5#6\rangle}

\newcommand{\xfo}{{x_\text{FO}}}

\newcommand{\lfs}{\lambda_\text{fs}}

\renewcommand{\sec}{\ensuremath{\mathrm{s}}}

\newcommand{\cm}{\ensuremath{\mathrm{cm}}}

\newcommand{\eV}{\ensuremath{\mathrm{eV}}}
\newcommand{\keV}{\ensuremath{\mathrm{keV}}}
\newcommand{\MeV}{\ensuremath{\mathrm{MeV}}}
\newcommand{\GeV}{\ensuremath{\mathrm{GeV}}}

\begin{document}
\begin{flushright}
PI/UAN-2017-600FT
\end{flushright}

\title{Simply split SIMPs}

\author[1, 2]{Nicolás Bernal,}
\emailAdd{nicolas.bernal@uan.edu.co}

\author[3]{Xiaoyong Chu}
\emailAdd{xiaoyong.chu@oeaw.ac.at}

\author[3]{and Josef Pradler}
\emailAdd{josef.pradler@oeaw.ac.at}

\affiliation[1]{\it Centro de Investigaciones, Universidad Antonio Nariño\\
Cra 3 Este \# 47A-15, Bogotá, Colombia}

\affiliation[2]{\it ICTP South American Institute of Fundamental Research\\
Instituto de Física Teórica, Universidade Estadual Paulista\\
R. Dr. Bento Teobaldo Ferraz 271, 01140-070 São Paulo, Brazil}

\affiliation[3]{\it Institute of High Energy Physics,\\
                Austrian Academy of Sciences, Nikolsdorfer Gasse 18, 1050 Vienna, Austria}

\abstract{
  Dark Matter which interacts strongly with itself, but only feebly
  with the Standard Model is a possibility that has been entertained
  to solve apparent small-scale structure problems that are pertinent
  to the non-interacting cold Dark Matter paradigm.
  In this paper, we study the simple case in which the self-scattering
  rate today is regulated by kinematics and/or the abundance ratio,
  through the mass-splitting of nearly degenerate pseudo-Dirac
  fermions $\chi_1$ and $\chi_2$ or real scalars $\phi_1$ and
  $\phi_2$.
  We calculate the relic density of these states in a scenario where
  self-scattering proceeds through off-diagonal couplings with a
  vector particle $V$ (Dark Photon) and where the abundance is set
  through number-depleting 4-to-2 reactions in the hidden sector, or,
  alternatively, via freeze-in.
  We study the implications of the considered models and their
  prospect of solving astrophysical small-scale structure problems. We
  also show how the introduction of the (meta-)stable heavier state
  may be probed in future dark matter searches.
}

\maketitle

\section{Introduction}
\label{sec:introduction}

The non-gravitational nature of Dark Matter (DM) remains a mystery.
Yet, the apparent similarities in the cosmic abundances of DM and
baryonic matter~\cite{Ade:2015xua}, $\Omega_\text{DM} \simeq 5\,\Omega_\text{B}$, may be considered
as indicative that both forms of matter were once in thermal
equilibrium with each other. If so, this demands a minimum strength in
coupling between the Standard Model (SM) and the hidden sector,
restricting severely the class of models that hosts a successful DM
candidate.
In the past, this class of models has received by far the biggest
attention, theoretically as well as experimentally.  Most prominent in
this class are extensions of SM that feature weakly interacting
massive particles (WIMPs) as DM. WIMPs carry electroweak-scale mass
and couple to SM with a strength that is reminiscent to that of the
weak interactions. The correct relic density is then achieved in the
chemical decoupling process from the SM. 

Once the link between the SM and DM is gradually severed, both sectors
may never reach thermal equilibrium with each other. The requirement
$\Omega_\text{DM} \simeq 5\,\Omega_\text{B}$ then becomes prone to initial
conditions and the details of the cosmic history, but the spectrum of
possibilities amplifies.  
One possibility is that the DM abundance is set entirely in the hidden
sector. This has been entertained recently in strongly
self-interacting DM models that annihilate in number-depleting
3-to-2~\cite{Dolgov:1980uu,Carlson:1992fn,Hochberg:2014dra,Yamanaka:2014pva,Hochberg:2014kqa,Bernal:2015bla,Bernal:2015lbl,Lee:2015gsa,Choi:2015bya,Hansen:2015yaa,Bernal:2015ova,Kuflik:2015isi,Hochberg:2015vrg,Choi:2016hid,Pappadopulo:2016pkp,Farina:2016llk,Choi:2016tkj,Dey:2016qgf}
or 4-to-2 interactions~\cite{Bernal:2015xba}: the so-called Strongly
Interacting Massive Particle (SIMP) mechanism, where ``strong'' must
not necessarily allude to a confining dark force.  Although chemical
equilibrium with the observable sector is never reached, DM may still
carry the imprint of the SM thermal bath through elastic scatterings
that keep DM in kinetic equilibrium with SM. In fact, the latter is
often imposed as a requirement to prevent DM particles in the final
state of the number-violating annihilation process to act as hot DM,
 significantly modifying structure formation~\cite{deLaix:1995vi}.
Finally, it is possible that the dark and observable sector never
reach kinetic equilibrium. It is the case adopted in this work.  The
``coldness'' of DM can then be ensured by allowing the initial DM
temperature $T'$ to be much smaller than the SM temperature
$T$~\cite{Bernal:2015ova,Bernal:2015xba,Heikinheimo:2016yds}.%
\footnote{Alternatively, one may consider an enlarged dark sector that
  contains additional relativistic states at the moment of the
  freeze-out~\cite{Bernal:2015bla,Bernal:2015lbl}, thereby preventing
  the heating of DM during freeze-out.}  Such initial condition can,
\emph{e.g.}, be dynamically achieved through a feeble coupling between
DM and SM particles, through a hierarchy in branching-ratios in
inflaton decay, or, more generally, by heating the SM sector through
the decays of some exotic particles.

Although laboratory signatures of DM are less certain in the latter
``non-WIMP''-type of models, 
elastic self-interactions of DM have consequences for structure
formation. Models of SIMP DM have therefore been entertained as a
solution to the small scale structure problems, which appear to
persist in the collisionless DM paradigm, such as the ``core vs.~cusp
problem''~\cite{Flores:1994gz,Moore:1994yx,Oh:2010mc,Walker:2011zu}
and the ``too-big-to-fail
problem''~\cite{BoylanKolchin:2011de,Garrison-Kimmel:2014vqa}.  These
can be alleviated if at the scale of dwarf galaxies there exists a large
self-scattering cross section, $\sigma$, over DM particle mass, $m$,
in the range
$0.1\lesssim\sigma/m\lesssim10$~cm$^2/$g~\cite{Spergel:1999mh,Wandelt:2000ad,Vogelsberger:2012ku,Rocha:2012jg,Peter:2012jh,Zavala:2012us,Vogelsberger:2014pda,Elbert:2014bma,Kaplinghat:2015aga}.
The self-scattering of DM particles leads to heat transfer that
decreases the density contrast in the centers of DM halos turning
cusps into cores and changing the subhalo abundance matching due to a
lower halo concentration. Self-interacting DM therefore directly
addresses the two small-scale
problems~\cite{Spergel:1999mh,Wandelt:2000ad,Vogelsberger:2012ku,Rocha:2012jg,Peter:2012jh,Zavala:2012us,Kaplinghat:2015aga},
while astrophysical solutions also
exist~\cite{MacLow:1998wv,Governato:2009bg, Silk:2010aw,
  VeraCiro:2012na}.  Although this effect alone cannot efficiently
reduce the formation rate of luminous galaxies in DM subhalos, it may
still alleviate the ``missing satellites
problem''~\cite{Klypin:1999uc,Moore:1999nt} with help of more DM
physics (\emph{e.g.} warm or decaying DM) or baryonic feedback.%
\footnote{While latest
  observations~\cite{Bechtol:2015cbp,Drlica-Wagner:2015ufc} tend to
  prefer the latter option, the problem remains unsettled~\cite{Koposov:2007ni,Jethwa:2016gra}.}
Finally, the non-observation of an offset between the mass
distribution of DM and hot baryonic gas in the Bullet Cluster
constrains the DM self-interaction cross section to $\sigma/m<1.25$
cm$^2$/g at
$68\%$~CL~\cite{Clowe:2003tk,Markevitch:2003at,Randall:2007ph},
\textit{i.e.}, approximately 1~barn %
for 1\,GeV DM mass. Similarly, recent observations of cluster
collisions lead to the constraint $\sigma/m<0.47-2$~cm$^2$/g at
$95\%$~CL~\cite{Harvey:2015hha, Robertson:2016qef}.

Here we consider the scenario that SIMPs come in the form of a finely
split mass doublet, where the ground state (state~1) and the heavier
state (state~2) share a common $\mathbb{Z}_2$ symmetry, stabilizing
state~1. Depending on the mass splitting between the two states, the
heavier one can be long-lived on cosmological time-scales.  
The scenario hence resembles ``inelastic
DM''~\cite{TuckerSmith:2001hy} where, at tree-level, each state only
scatters with its counterpart, but not with itself. As a result, the
effect of DM self-scattering depends on the abundances of both states.
This provides a new way to regulate the DM self-interaction,
alleviating strong astrophysical constraints in certain cases.  In
practice, it is achieved by modifying either the DM annihilation rate
or the lifetime of heavier state. In the case of a long-lived heavier
state, it leads to a two-component DM model. In order to achieve
strong enough self-interaction, we are concerned with nearly
degenerate states with sub-GeV DM mass, hence distinct from both
exothermic double-disk~\cite{McCullough:2013jma} and boosted
DM~\cite{Agashe:2014yua}.  Moreover, the existence of the heavier
state in our scenario leaves distinguishable imprints in the
low-redshift Universe, as well as in DM search experiments.

It is the purpose of this work to investigate the viability of the
above scenario for both scalar and fermion DM. In this scenario  the dark and observable sectors carry distinct
temperatures. Such difference in temperatures can \textit{e.g.}~be
generated in a small branching from inflaton decay into DM
~\cite{Dev:2013yza,Kane:2015qea}. The dark sector then reached
chemical equilibrium within its own sector, while remaining decoupled
from the SM sector.  Alternatively, a difference in temperatures could
also be dynamically generated, via a small portal between DM and SM
particles through freeze-in~\cite{Hall:2009bx,McDonald:2001vt,Bernal:2015xba}.
For maintaining a nearly decoupled dark sector, we assume that the
dark particles only feebly couple to SM particles, via a vector portal
coupling. Throughout the study, the vector $V$ mediating the DM
self-interaction is assumed to be heavier, so that Sommerfeld
enhancement plays no role~\cite{Slatyer:2009vg, Schutz:2014nka,
  Zhang:2016dck}. Note that the $\mathbb{Z}_2$ symmetry between two
states prohibits 3-to-2 self-interactions, but instead allows for
4-to-2 annihilation (or, alternatively, freeze-in) to generate the
observed DM density.

The paper is organized as follows. In Sec.~\ref{sec:simple-models} we
introduce simple models for finely split states.  In
Sec.~\ref{sec:dark-matter-relic} we discuss the generation of the DM
relic abundance via the 4-to-2 annihilations (or freeze-in) for two
scenarios: pseudo-Dirac and real scalar DM.  Sec.~\ref{sec:astro}
is devoted to the astrophysical implications of split SIMPs; searches
for possible connections between the dark and the visible sectors are
discussed in Sec.~\ref{sec:connection}.  The conclusions are presented
in Sec.~\ref{sec:conclusions}.  Some formul\ae\ used in this work are
provided in the Appendices. %

\section{Simple Models with Off-diagonal Interactions}
\label{sec:simple-models}

A natural and simple possibility to ensure the dominance of inelastic
interactions in self-scatterings is through the %
mediation of a
massive vector particle $V_{\mu}$ when the components of a Dirac
fermion $\Psi$ are split by small Majorana masses $m_L,\ m_R$, or the
real and imaginary parts of a complex scalar $\Phi$  by a
mass-squared parameter $m^2_{\phi}$,
\begin{align}
  \mathcal{L}_{\Psi} & = \bar \Psi \left( i \slashed D - M_D \right) \Psi 
  - \frac{m_L}{2} \left( \bar\Psi^c P_L \Psi + h.c. \right) - 
    \frac{m_R}{2} \left( \bar\Psi^c P_R \Psi + h.c. \right) ,\\
    \mathcal{L}_{\Phi} & = |D_{\mu} \Phi|^2 + M^2 |\Phi|^2 + (m_{\phi}^2\,\Phi^2 + h.c.) - V_{\Phi}\,. 
\end{align} 
Here, $D_\mu\equiv \partial_\mu  + i g_V V_\mu$ is the covariant derivative with
gauge coupling $g_V$, $\Psi^c = \mathcal{C} \bar\Psi^T $ denotes the
charge conjugate state, $P_{R,\,L}\equiv \frac{1}{2}(1\pm \gamma^5)$ are chirality projectors and $H$ is the SM Higgs doublet. The
scalar potential of $\Phi$ is given by
\begin{align}
  V_{\Phi} =  \lambda_\Phi |\Phi|^4 + \frac{\lambda'_\Phi }{2} ( \Phi^4 + h.c. )+  
  \lambda_m |\Phi|^2 |H|^2 +  
  (\lambda'_{m} \Phi^2 + h.c. ) |H|^2 \,. 
\end{align}
As we are interested in the possibility of split DM, the quartic self-couplings, $\lambda_\Phi$ and $\lambda'_\Phi$, are assumed to be negligible unless otherwise stated.

We assume $m_{L,\,R} \ll M_D$ and $m_{\phi} \ll M$ for which $\Psi$
decomposes into two mass-diagonal Majorana states
$\chi_{1,\,2} = \chi_{1,\,2}^c $ and $\Phi$ into two real scalar fields
$\phi_{1,\,2}$, 
\begin{align}
 \chi_1 &\simeq \frac{i}{\sqrt{2}} (\Psi - \Psi^c)\,,\quad
 \chi_2 \simeq \frac{1}{\sqrt{2}} (\Psi + \Psi^c)\,,\quad  \Phi = \frac{1}{\sqrt{2}}(\phi_1 + i\, \phi_2)\,. 
\end{align}
For the fermion states $\chi_{1,\,2}$ %
the mixing angle
that diagonalizes the mass matrix is maximal ($\pi/4$) up to
corrections $\delta\equiv (m_L -m_R)/(2M_D) \ll 1$ .  The masses are given by
\begin{align}
 \text{Pseudo-Dirac }\chi_{1,\,2}:\quad   m_{1,\,2} &\simeq M_D \mp  \frac{m_L + m_R}{2} + O(\delta)\,,\\
 \text{Scalar }\phi_{1,\,2}:\quad  m_{1,\,2}^2 & = M^2 \mp  m_{\phi}^2\,.
\end{align}
We note in passing that a complex $m_\phi$ would not lead to a similar
$O(\delta)$ term in the scalar case, and we set all mass parameters to
be real.
In the following it will be convenient to parametrize the model in
terms of $m \equiv m_1$ and $\Delta m \equiv m_2 - m_1 > 0$.  The
interactions in the mass eigenbasis read
\begin{align}
  \mathcal{L}_{\rm int,\, \chi} &  =
                                i g_V \bar\chi_1 \gamma^{\mu} \chi_2 V_{\mu} + \frac{g_V }{2}  \delta\left( \bar\chi_1 \gamma^{\mu}\gamma^5 \chi_1  -  \bar\chi_2 \gamma^{\mu}\gamma^5 \chi_2   \right) V_{\mu}\,, \\
  \mathcal{L}_{\rm int,\, \phi} & = g_V \left(\phi_1 \partial^{\mu} \phi_2 - \phi_2 \partial_{\mu} \phi_1 \right)V_{\mu} + \frac{1}{2} g_V^2 \left(\phi_1^2 + \phi_2^2\right) V^2 - V_\Phi(\phi_1,\,\phi_2)\,. 
\end{align} 

Finally, the Lagrangian for the new gauge boson is given by
\begin{align}
  \mathcal{L}_V = -\frac{1}{4}V_{\mu\nu}V^{\mu\nu}  + \frac{m_V^2}{2} V^2  - \kappa\,V_{\mu} J^{\mu}_{\rm SM}\,, 
\end{align}
where in the last term we consider an effective coupling to a SM
current $J^{\mu}_{\rm SM} $. A particular prominent choice is the
gauge kinetic mixing~\cite{Holdom:1985ag} of the $V$ and hypercharge
($Y$) field strengths,
$-\kappa/(2\cos\theta_W)\,F_{\mu\nu}^Y V^{\mu\nu}$. In the low-energy
effective theory, the coupling to the electromagnetic current is most
important and hence
$J^{\mu}_{\rm SM} = e \sum_f q_f \bar f \gamma^{\mu} f$ above, with
$e$ being the electromagnetic gauge coupling and $q_f$ the charge of
SM fermion~$f$.  The coupling of $V$ to the SM neutral current is
suppressed by $m_V^2/m_Z^2$ and thus negligible for $m_V \ll m_Z$.  We
hence refer to $V$ as ``dark photon'' throughout the paper.  In all
cases, we also impose the condition $m_V > m_1+m_2$ to ensure
efficient decay of $V$ within the dark sector.

Before moving to the next section, we comment on diagonal interactions
that may appear.  On the one hand, in the fermionic case the mass
splitting is typically generated by a vacuum expectation value
of a doubly dark-charged scalar $S$ with the interaction term
$S \left(\bar \Psi^c\,\Psi\right)$, that leads to a diagonal
interaction term proportional to the mass splitting $\Delta m$,
\textit{i.e.} $m_L + m_R$. For $\Delta m \ll m$, its contribution to
the DM elastic scattering is typically much smaller than the one
produced by box diagrams with the exchange of two $V$ bosons.  A
similar conclusion holds for interactions proportional to $\delta$,
\textit{i.e.}, for the parity-violating difference between $m_L$
  and $m_R$. On the other hand, for scalar DM, the quartic couplings
  $\lambda_{\Phi}$, $\lambda_{\Phi'}>0$ in the scalar potential
  $V_{\Phi}$ yield diagonal interactions.  Throughout this work, we
  require them to be weaker than the ones induced by exchanging two
  $V$ bosons (see below). This holds true when
\begin{equation}
\lambda_{\Phi}+\lambda_{\Phi}' 
 \le \, {\mathcal O}\left(10^{-3} \right)\,g_V^4  \,\left(\frac{2.5\,m}{m_V}\right)^4.
\end{equation}
Likewise, we assume $\delta = 0$, so that diagonal interactions do not
play a role in either DM self-scattering or DM 4-to-2 annihilation.

\section{Dark Matter Relic Abundance}
\label{sec:dark-matter-relic}

We first study the generation mechanisms of the observed DM density,
and how the relative abundance of states 1 and 2 is determined. Within
the framework of standard cosmology, the final abundance of DM
particles can be generated either via 4-to-2 annihilations or via
freeze-in. We start with the former and discuss the latter towards the
end of this section.

We assume that the dark sector (states 1, 2 and $V$) was once in
equilibrium with each other, so that we can define the relative
abundance of states 1 and 2 by their corresponding ratio
\begin{equation}
\label{eq:R}
  R(T')\equiv\frac{n_2(T')}{n_1(T')} =\left(1 + \frac{\Delta m}{m}   \right)^{3/2}e^{-\frac{\Delta m}{T'}} \simeq e^{-\frac{\Delta m}{T'}}\,. 
\end{equation}
Here $T'$ denotes the dark sector temperature which is, in general,
different from the photon temperature~$T$.  Eq.~\eqref{eq:R} is valid
when both states are non-relativistic and share a common chemical
potential, \textit{i.e.}, while the $11\leftrightarrow22$ reaction is
in equilibrium.  When considering finely split states $\chi_{1,\,2}$
or $\phi_{1,\,2}$ it is natural to solve for the sum of the respective
number densities $n\equiv n_1 + n_2 = n_1\,(1+R)$ in the relic density
calculation.  Let us point out that the $11\leftrightarrow22$ reaction
does not affect the total DM number density; nevertheless, the latter
process is crucial to determine the value of $R$ at late times.

\begin{figure}[tb]
  \centering
\includegraphics[width=0.85\textwidth]{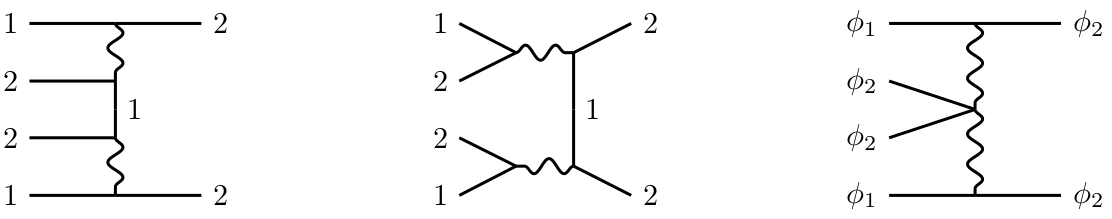}
\caption{\small Exemplary 4-to-2 annihilation diagrams.  The first two
  diagrams (left and central panels) are the only non-vanishing
  processes in the leading non-relativistic limit for pseudo-Dirac DM.
  For scalar DM there is a another topology (right panel) which
  involves a quartic gauge coupling that is diagonal in $\phi_i$.  }
  \label{fig:fourtoone}
\end{figure}

In the SIMP mechanism, the final DM abundance is fixed by
number-depleting processes, some of which are shown in
Fig.~\ref{fig:fourtoone}.  Quantitatively, the evolution of the
dark-sector comoving number density is governed by the Boltzmann
equation
\begin{equation}
\label{eq:Y}
\frac{dY}{dx}=-\frac{s^3\,\langle\sigma v^3\rangle_{4\to2}}{x\,H}\left(Y^4-Y^2\,Y_\text{eq}^2\right)\,,
\end{equation}
where $x\equiv m/T$ and $Y\equiv n/s$ is the yield variable, including both states 1 and 2, defined in
terms of the SM entropy density $s(T)$. $Y_{\rm eq}$
denotes the equilibrium yield in the dark sector.
The thermally averaged cross section is given by
\begin{eqnarray}
\label{eq:sigv}
\langle\sigma v^3\rangle_{4\to2}
 &\equiv&\frac{1}{(1+R)^4}\Big[\NN{2}{2}{2}{2}{1}{1}R^4+\NN{1}{2}{2}{2}{1}{2}R^3+\big(\NN{1}{1}{2}{2}{2}{2}+\NN{1}{1}{2}{2}{1}{1}\big)R^2\nonumber\\
&&\qquad\qquad+\NN{1}{1}{1}{2}{1}{2}R+\NN{1}{1}{1}{1}{2}{2}\Big]\,.
\end{eqnarray}
The definition of the expressions $\NN{i}{j}{k}{l}{a}{b} $ is detailed
in Appendix~\ref{sec:annihilation-rates}. Since $Y$ describes the total
abundance of both states 1 and 2, only total number-changing processes
are taken into account, including co-annihilation channels. In this
expression, the entropy density $s$ and the Hubble rate $H$ are
functions of the temperature $T$, whereas $R$ is a function of $T'$
(or equivalently $x'\equiv m/T'$).
In our study, we solve the Boltzmann equation~\eqref{eq:Y} numerically
and obtain the freeze-out time and the DM relic abundance (see Appendix~\ref{app:boltzmann}). %

As already pointed out in Ref.~\cite{Carlson:1992fn}, if dark and
  visible sectors are decoupled, one can equivalently track the
evolution %
from the separate conservation of respective comoving entropies $S'$
and $S$. The DM abundance at freeze-out can then be estimated
as~\cite{Bernal:2015xba}
\begin{equation}
\Omega_\text{DM} h^2 \sim \frac{m}{3.6~\text{eV}(x'_\text{FO} +2.5)} \frac{S'}{S}\,, \label{entropy:ratio}
\end{equation}
which in turn allows to determine the entropy ratio,
from the observed DM relic abundance and for a given freeze-out point, $x'_\text{FO}$.
In practice, the value of $x'_\text{FO}$ is of order
${\mathcal O}(15)$ (see left panel of Fig.~\ref{fig:lfs}), so for each DM mass the
final relic abundance is largely fixed by the entropy ratio, and vice versa.

Importantly, the fact that $x'_\text{FO}$ cannot be arbitrarily large
yields an upper bound on the entropy ratio, and thus on total extra
energy density contributed by the dark sector. The latter is often
quantified in the effective number of relativistic neutrino degrees of
freedom, $ N^\nu_{\rm eff} = 3.046 + \Delta N^\nu_{\rm eff} $ where
$\Delta N^\nu_{\rm eff}$ measures beyond SM contributions.
Quantitatively, Eq.~\eqref{entropy:ratio} together with separate
entropy conservation suggest that
\begin{equation}
\label{eq:Neff}
  \left.\Delta  N^\nu_{\rm eff} \right|_{\rm BBN} \lesssim 10^{-2} \,
  \left[\frac{x'_\text{FO} +2.5}{16}\,\frac{{\rm keV}}{m}\right]^\frac43
\end{equation} 
at $T\sim 1$\,MeV, assuming a relativistic dark sector at that
moment. %
From Eq.~\eqref{eq:Neff} it follows that for DM heavier than a few keV,
the dark sector does not contribute to extra radiation in quantities
where it is currently constrained: the concordance of Big Bang
Nucleosynthesis (BBN) predictions and observationally inferred
primordial light element abundances of D and ${}^4$He imply
$ N^\nu_{\rm eff} = 2.88 \pm 0.16 $~\cite{Cyburt:2015mya}
(marginalized over the CMB determined baryon density). The scenario is
also consistent with the CMB determination without BBN,
$N^\nu_{\rm eff} = 3.15 \pm 0.23 $~\cite{Ade:2015xua}, because dark
particles will have become non-relativistic at matter-radiation
equality for $m\gtrsim 1~\keV$.

\subsection{Pseudo-Dirac Dark Matter}

\begin{figure}[t!]
\centering
\includegraphics[width=0.49\textwidth]{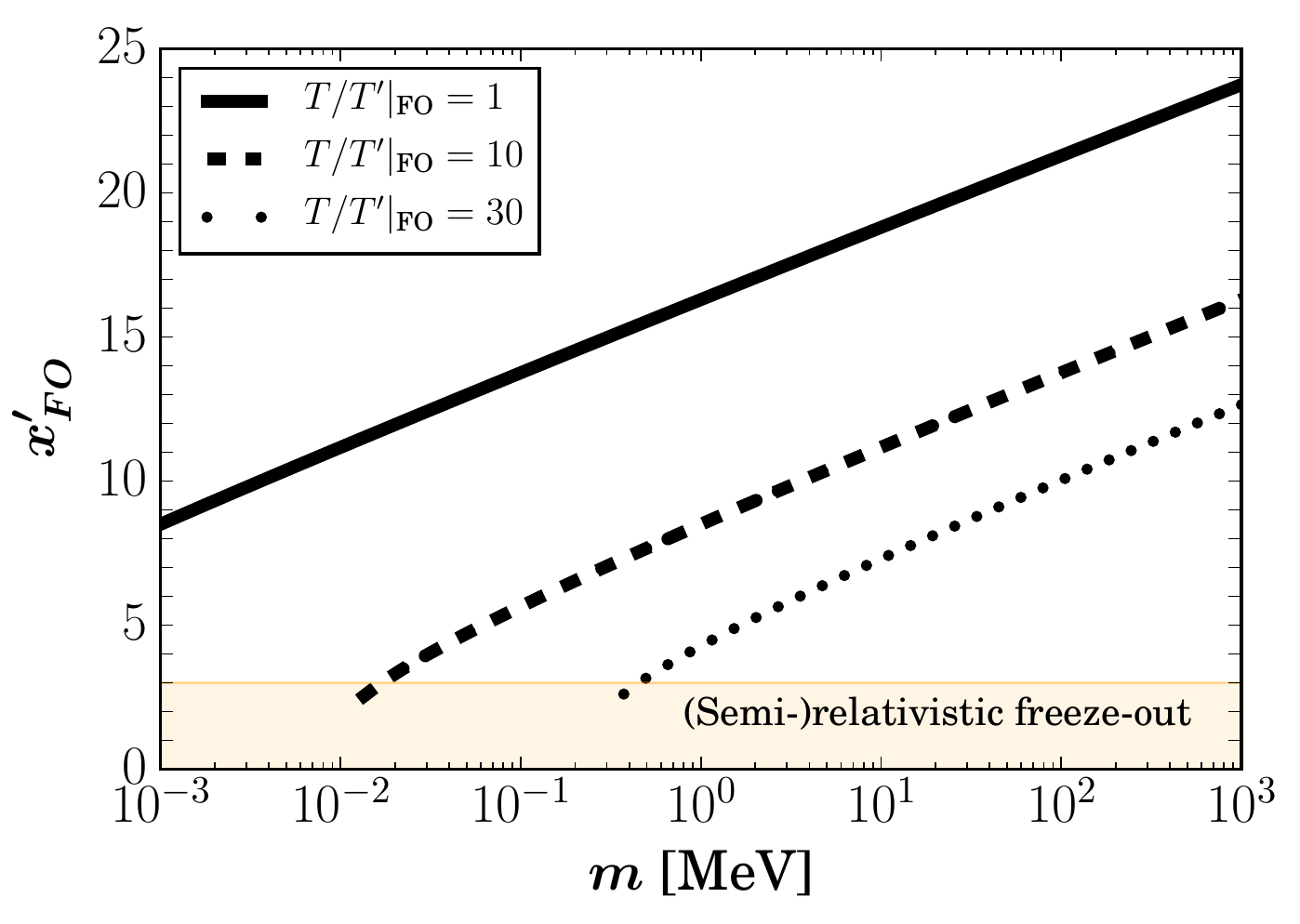}
\includegraphics[width=0.49\textwidth]{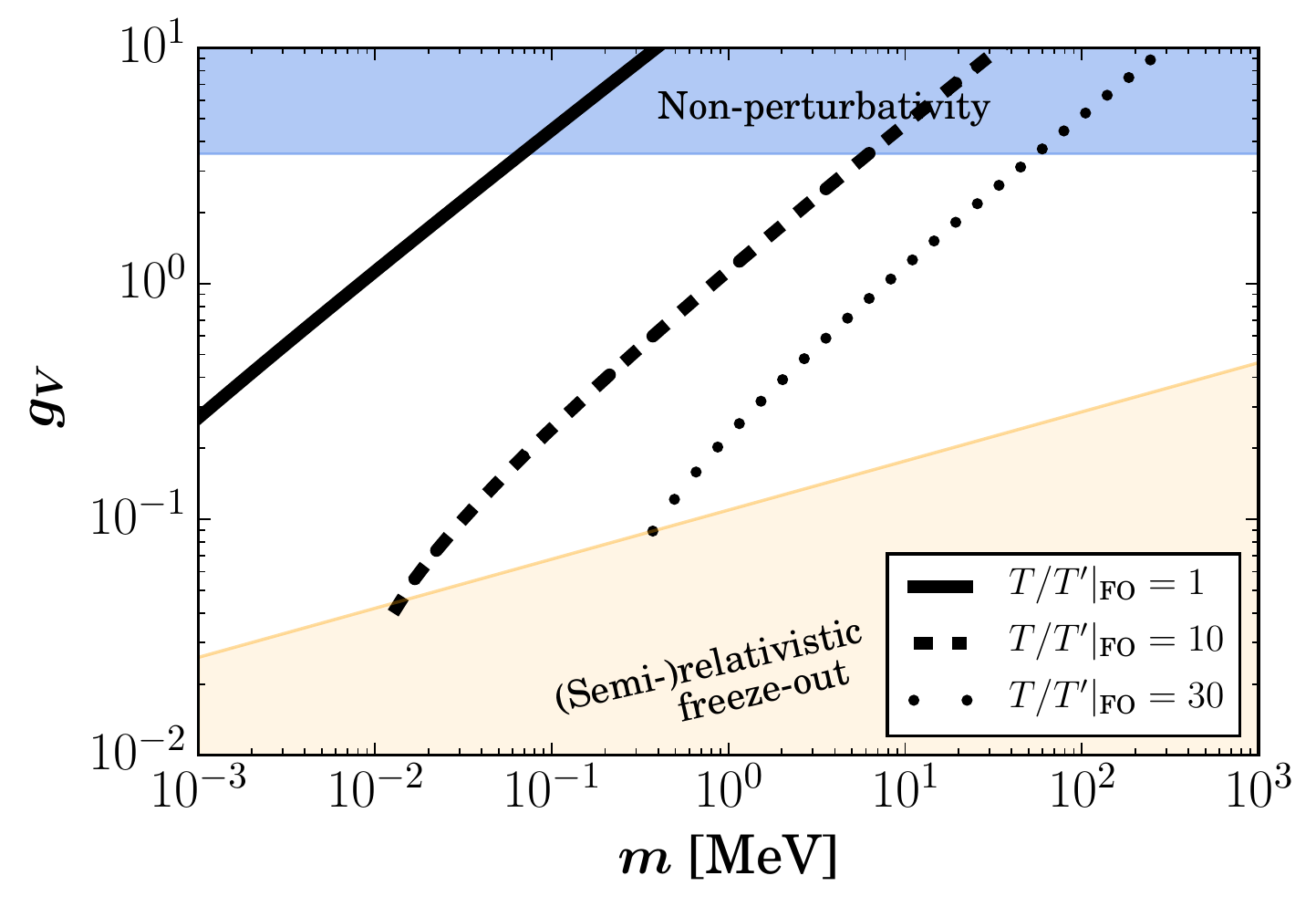}
\caption{Pseudo-Dirac DM freeze-out in the non-relativistic regime as
  a function of the DM mass $m$, for $\Delta m/m =10^{-2}$ and various
  temperature ratios at DM freeze-out. The \emph{left} panel shows the
  freeze-out point $x'_\text{FO}$ that yields the observed DM density,
  while the \emph{right} panel shows the corresponding value of the
  dark gauge coupling $g_V$, assuming $m_V = 2.5~m$.}
\label{fig:lfs}
\end{figure}
For fermionic DM, the only non-vanishing S-wave annihilation process
is $2\chi_i + 2\chi_j \to 2\chi_i$ with $i\neq j$, due to the Pauli
exclusion principle.%
\footnote{Here the term ``S-wave'' refers to vanishing angular
    momentum between any pair of the incoming particles.}
That is, the 4-to-2 interaction rate, Eq.~\eqref{eq:sigv}, becomes
\begin{equation}
\langle\sigma v^3\rangle_{4\to2}= \big[\NN{1}{1}{2}{2}{2}{2}+\NN{1}{1}{2}{2}{1}{1}\big]\frac{R^2}{(1+R)^4}\,,
\end{equation}
being suppressed in the non-relativistic regime by the number density
of state~2.  In the previous equation, the two relevant annihilation
cross sections are given by
\begin{equation}
\langle1122\to11\rangle=\langle1122\to22\rangle=\frac{27\sqrt{3}\,\gd^8}{32\pi}\frac{\left(\mzp^4-8\mx^2\mzp^2-8\mx^4\right)^2}{\left(\mzp^4-2\mx^2\mzp^2-8\mx^4\right)^4}\,,
\end{equation} 
up to corrections proportional to $\Delta m$.  Typical diagrams
contributing to the process are shown in the left and middle panels of
Fig.~\ref{fig:fourtoone}. In the middle panel, if $m_V=m_1+m_2$ both
vector propagators can simultaneously go on-shell, substantially
enhancing the efficiency of the process. In our analysis, this resonance
is never reached since $m_V > m_1+m_2$ has been
imposed in order to have a rapid decay of $V$ into states~1 and~2.

Figure~\ref{fig:lfs} presents different aspects of the solution of the
Boltzmann equation~\eqref{eq:Y} in the non-relativistic regime: for
several temperature ratios $T/T'$ at
freeze-out, the figure shows the freeze-out point (left panel) and the dark gauge coupling
$g_V$ (right panel) for achieving the measure DM abundance, as a function of the DM mass.
The upper blue region, corresponding to $g_V>\sqrt{4\pi}$, goes beyond our perturbative approach.
It is shown that for an increasing DM mass, a larger $x'_\text{FO}$
(and hence a stronger Boltzmann suppression) is required in order to
achieve the correct relic density.  This implies a later freeze-out
and in turn a stronger gauge interaction.  A suppression of the %
DM yield can also be achieved by imposing a colder dark
sector, where smaller values of $x'_\text{FO}$ and $g_V$ can produce
the observed DM density.
The right panel suggests the ballpark of interest for DM masses:
for $m\gtrsim \GeV$ one faces the perturbativity limit; in turn,
fermionic DM is also bounded from below, $m\gtrsim 1\,\keV$, by the
Gunn-Tremaine limit~\cite{Tremaine:1979we,Boyarsky:2008ju}. For
the remainder of this work we restrict ourselves to the keV-GeV mass
bracket.
In the lower regions of both panels, freeze-out proceeds when DM is
(semi-)relativistic ($x'_\text{FO}<3$). There, the final DM abundance
can be directly obtained from the temperature ratio, and independently
of the gauge coupling $g_V$~\cite{Sigurdson:2009uz}. %
\begin{figure}[t!]
\centering
\includegraphics[width=0.49\textwidth]{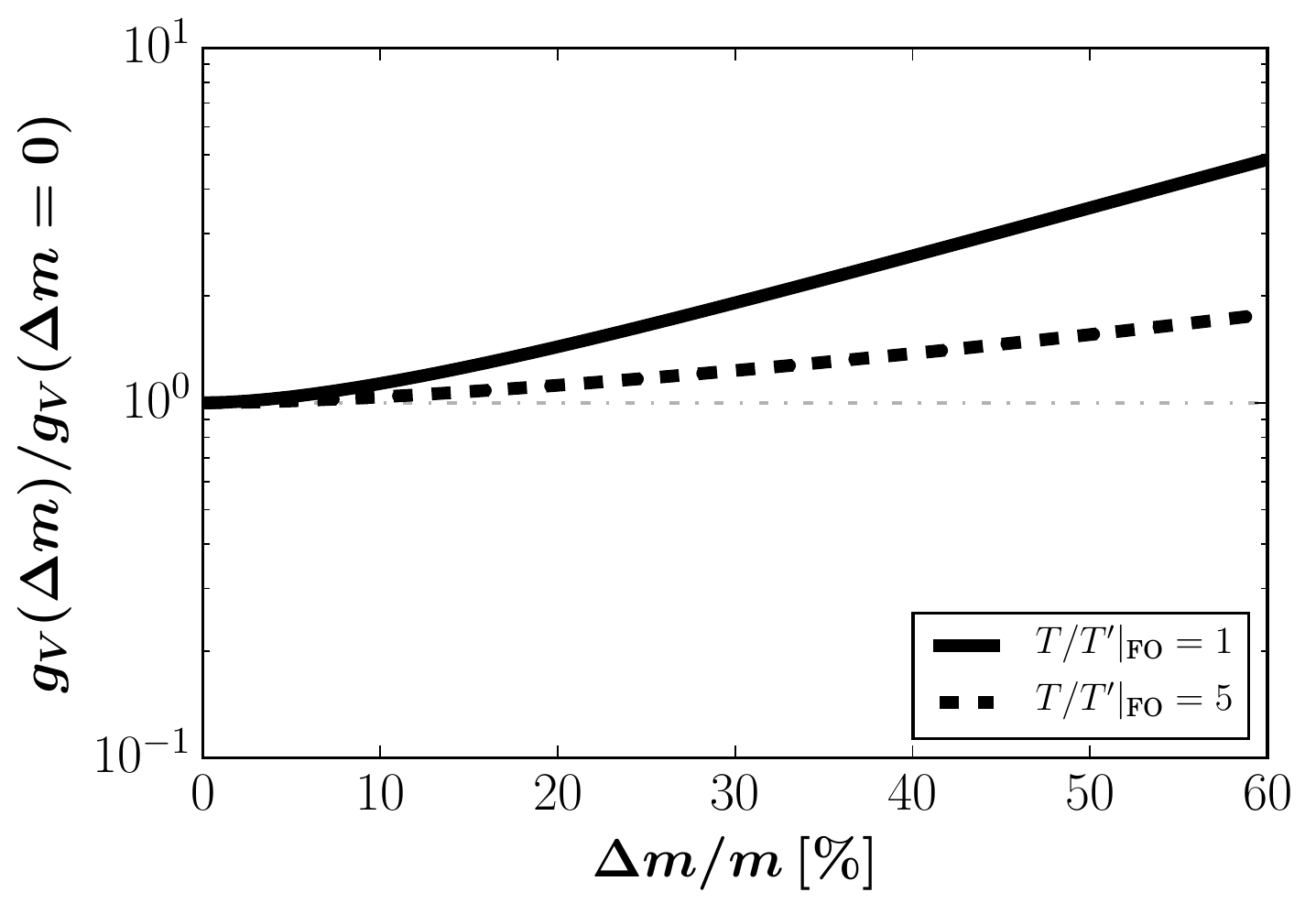}
\includegraphics[width=0.49\textwidth]{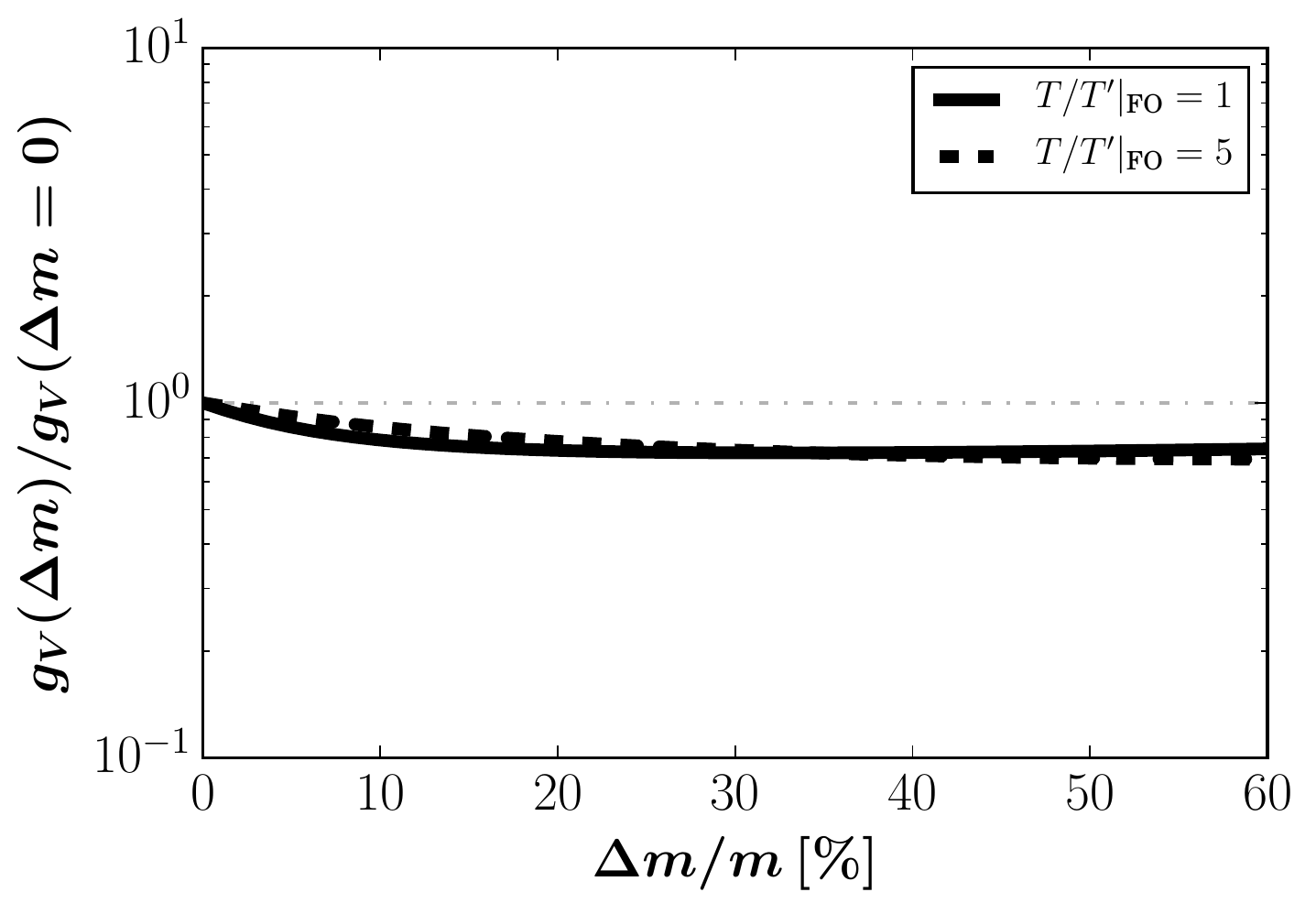}
\caption{Gauge coupling $g_V$, derived from the observed relic DM
  abundance, as a function of the mass difference $\Delta m/m$ for
  several fixed values of temperature ratios at freeze-out,
  $T'/T|_{FO}$. The left (right) panel is for fermion (scalar) DM with
  $m=10$~keV and $m_V = \frac52 m +\Delta m$; only terms constant
    in relative velocities are considered.}
\label{scalar:fermion}
\end{figure}

The left panel of Fig.~\ref{fig:lfs} also shows that $x'_\text{FO}$ is
much smaller than the assumed value $(\Delta m/m)^{-1} = 100$ for the
entire parameter region.  That is, the number density of state 2
remains relatively unsuppressed at freeze-out, which guarantees that S-wave
annihilation dominates. This can also be seen from the left panel of
Fig.~\ref{scalar:fermion}, where we plot the values of $g_V$ needed in
order to produce the observed DM relic abundance through S-wave
annihilation for fermionic DM. For the plot, we choose $m=10$~keV and
$m_V=\frac52 m+\Delta m$. It illustrates how the increase of
$\Delta m/m $ leads to a smaller S-wave annihilation rate, which in
turn needs to be compensated by a larger $g_V$. Such a feature is more
noticeable if the freeze-out happens later, \textit{i.e.},  larger $x'_{\rm FO}$, corresponding to  higher
masses or smaller $T/T'|_\text{FO}$ (see left panel of Fig.~\ref{fig:lfs}). 
This can be understood from Eq.~\eqref{eq:R}: larger $x'_{\rm FO}$  leads to stronger Boltzmann
suppression in $R$.  In practice, velocity suppressed contributions to
annihilation do not become important until $\Delta m/m \ge 10\%$
($20\%$) for $T/T'|_{\rm FO} = 1 $ ($5$).  Hence, for our purposes, it
is safe to neglect them, as we will be mostly concerned with mass
splittings $\Delta m/m \lesssim 1\%$ in the following sections, which
are phenomenologically more interesting.

\subsection{Real Scalar Dark Matter}

In a similar fashion as for pseudo-Dirac DM we obtain the solution to
the Boltzmann equation for the two finely split real scalars $\phi_{1,\,2}$.
The major difference is the absence of Pauli blocking in the initial
state of four scalars; channels involving more than two identical
particles in the initial state, such as
$3\phi_i+\phi_j\to\phi_i+\phi_j$ and $4\phi_i\to2\phi_j$, with
$i\ne j$, are now allowed. The cross
sections read
\begin{align}
  \langle ijkl\to mn\rangle \simeq  a_{ijkl}\,  \frac{9 \sqrt{3}}{32\pi}\frac{g_V^8}{(2m^2+ m_V^2)^4}\,,
\end{align}
where $a_{1122} = 1 $, $a_{1111} = a_{2222} = 12$ and $a_{1112} =a_{1222} = 4$. %
As a result, the freeze-out process is not
suppressed by either the number density ratio $R$ or the relative velocity, \textit{i.e.}, all the
terms in Eq.~\eqref{eq:sigv} are present in this case.

The dependence of $g_V$ on $\Delta m/m$ for scalar DM freeze-out,
using the same DM and vector masses as before, is illustrated in the right
panel of Fig.~\ref{scalar:fermion}. As can be seen, the required
values of $g_V$ to obtain the DM abundance is less sensitive to
the mass difference for scalar DM. Once $\Delta m/m \gtrsim 20\%$
  annihilation via $1111\to22$ becomes dominant while others become
  suppressed by $R$. Since the interaction rate of $1111\to22$ is
  approximately independent of the value of $\Delta m/m$ at
  freeze-out, so is $g_V$.

\subsection{Relative Abundance of the States 1 and 2 at the Decoupling}

In this subsection we discuss the abundance ratio $R$ defined in
Eq.~\eqref{eq:R}, and especially its value $R_{\rm dec } \equiv R(T'_{\rm dec})$
after the decoupling of the reaction $22\leftrightarrow11$ at a dark
sector temperature $T'_{\rm dec}$.  Its value today, $R_0$, will be
discussed in Sec.~\ref{sec:decaying-dm-after}; astrophysical
implications in the low-redshift Universe depend critically on the
latter.
It is clear that $R_0 = 0$ if the decay rate of state 2 satisfies
$\Gamma_2 \gg H_0$, where $H_0$ is the Hubble
 constant. %
However, if state 2 is meta-stable on cosmological timescales,
$\Gamma_2 \ll H_0$, then $R_0\simeq R_{\rm dec}$.

In the SIMP mechanism studied above, where dark sector thermalizes in
the early Universe, $T'_{\rm dec}$ is found from
\begin{equation}\label{eq:de}
n_2(T'_\text{dec}) \langle \sigma_\text{22$\to$11} v\rangle = n(T'_\text{dec})  \left[\frac{R(T'_\text{dec})}{1+ R(T'_\text{dec})} 
 \langle \sigma_\text{22$\to$11} v\rangle\right ]= H(T_\text{dec})\,, 
\end{equation}
where $n(T'_\text{de})$ is fixed by the observed DM relic abundance.
Note that the relation between $T'_\text{dec}$ and $T_\text{dec}$, or
equivalently, between the entropies of the two sectors, is given by
the solution of the Boltzmann equation~\eqref{eq:Y}, as mentioned
above.  Due to its large cross section, the kinetic decoupling
typically happens after the 4-to-2 freeze-out.
Although it is difficult to solve the above equation analytically,
$R(T'_\text{dec})$ may be estimated as follows: given that
$n(T'_\text{dec})$ is fixed by the observed DM abundance, the factor
inside the square bracket in Eq.~\eqref{eq:de}  should be close to the thermal annihilation
cross section of ordinary WIMP DM at decoupling. That is,
$R_{\rm dec} \sim {{\mathcal O}\text{(1)~pb} /\langle
  \sigma_\text{22$\to$11} v\rangle}$. Nevertheless, we caution the
reader that this analogy fails once $m/T'_\text{dec} \gg 25$, or, in
practice, when $\Delta m/m \lesssim 10^{-2}$, and in the
  following we solve Eq.~\eqref{eq:de} numerically.

\begin{figure}[t]
\centering
\includegraphics[width=0.49\textwidth]{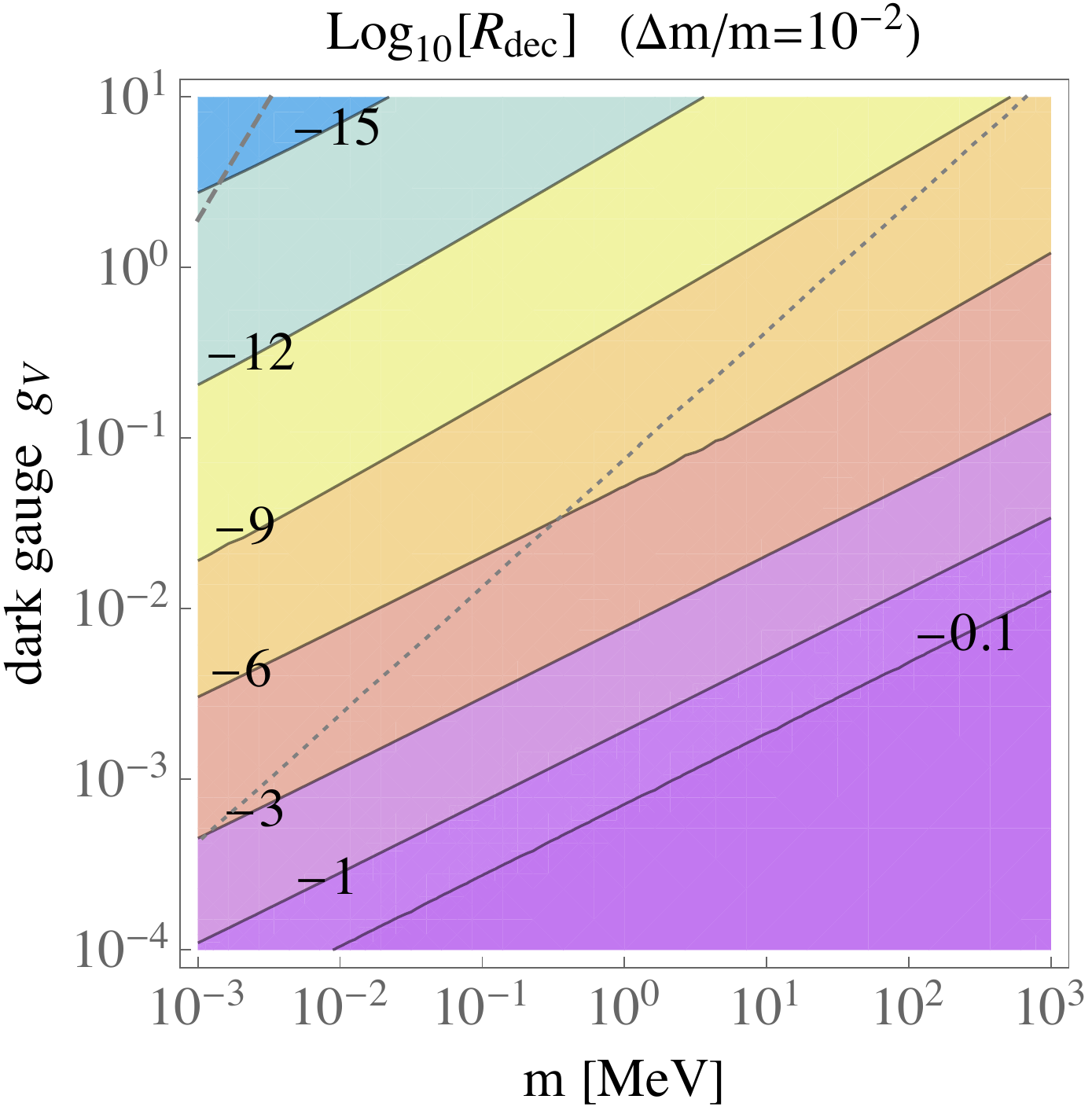}
\includegraphics[width=0.49\textwidth]{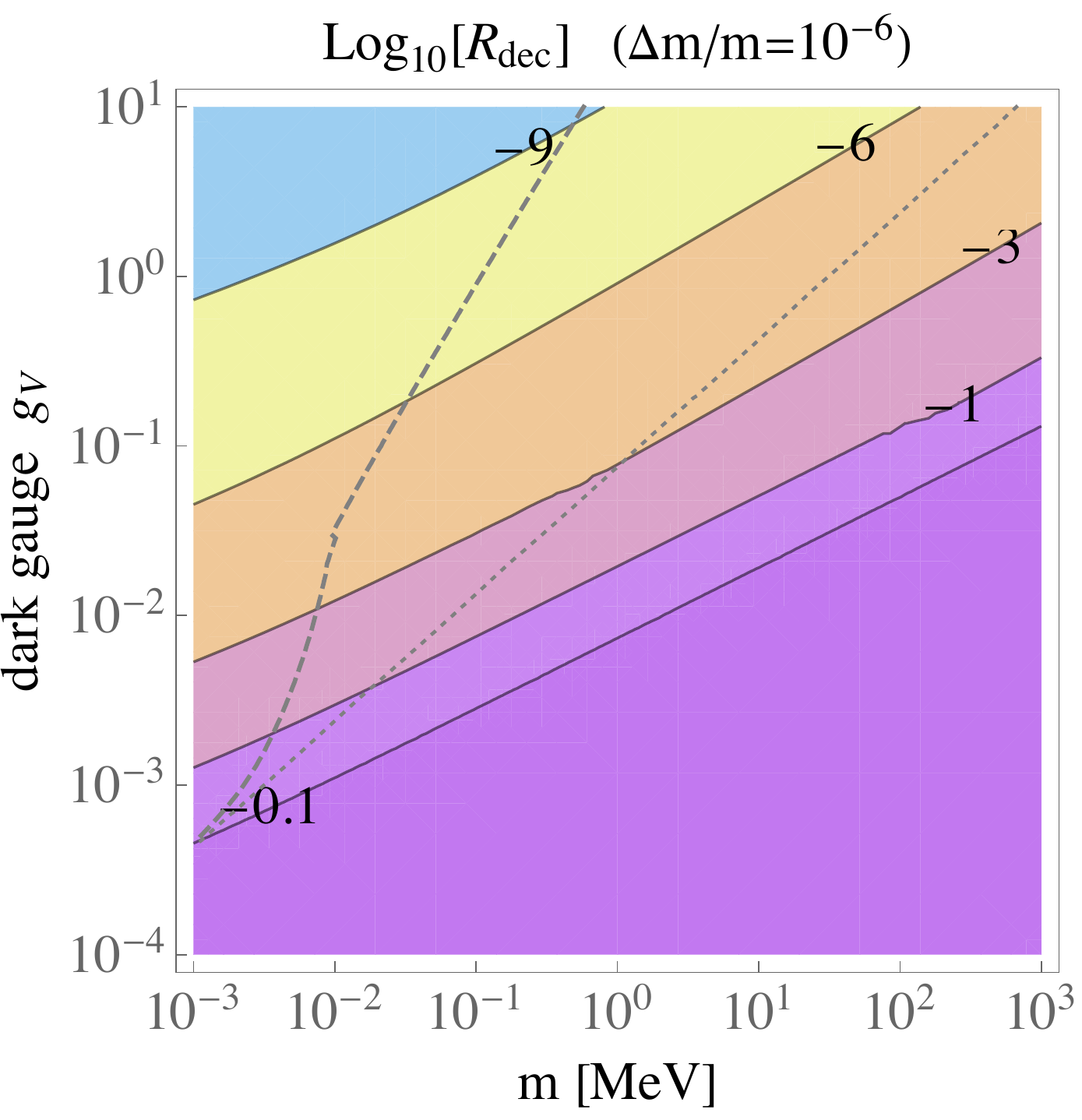}
\caption{ Contour lines for Log$_{10} R_{\rm dec}$ for fermionic DM
  with $\Delta m/m = 10^{-2}$ (left panel) and $10^{-6}$ (right
  panel), and $m_V = \frac52 m$.  In the region below the contour
  $-0.1$, $n_1 \simeq n_2$ approximately holds.  In each panel, dashed
  and dotted lines show $R_0\,\sigma_{12}/m = 1$~cm$^2$/g and
  $\sigma_{12}/m = 1$~cm$^2$/g, respectively, assuming
  $R_0 = R_{\rm dec}$ (explained below).  All the points reproduce the
  observed DM relic abundance.  }
\label{R0:msplit}
\end{figure}

Figure~\ref{R0:msplit} shows the values of $R_{\rm dec}$ that result
from fixing the freeze-out abundance to the observed DM relic density
(by adjusting $T/T'|_{\rm FO}$) for two different mass splittings:
$\Delta m/m = 10^{-2}$ (left panel) and $10^{-6}$ (right panel).  A
large mass splitting results in a negligible abundance of state~2.  On
similar grounds, large values of $R_{\rm dec}$ are observed in the
region of small $g_V$ and large $m$. The latter combination
corresponds to regions of small annihilation cross section, for which
an appreciable number of particles of type 2 must be present to
facilitate the 4-to-2 processes.  $R_{\rm dec} \lesssim 1$ is only
achieved when $T'_\text{dec}\ge \Delta m$ and together with
Eq.~\eqref{eq:de} this implies
\begin{equation}
g_V \lesssim 3\times 10^{-4}   \frac{m_V}{\sqrt{m \, \text{MeV}}}  
\left(\frac{ m^2}{\Delta m^2 } \frac{m\,T'_\text{FO}}{T^2_\text{FO}} \right)^\frac18\,,
\label{relaFO:R}
\end{equation}
where, again, the observed DM abundance has been used as an input.

Before ending this section, we comment on the case of the freeze-in
via the production of $V$ from SM.%
\footnote{Freeze-in of on-shell dark photons has, \textit{e.g.}, been
  calculated in detail in Ref.~\cite{Fradette:2014sza}.}
If the DM abundance is generated by freeze-in, the DM number density
is proportional to the portal interaction, as the latter transfers
energy from the SM thermal bath to the dark sector.%
\footnote{Re-annihilation, as a mixture of freeze-in and
  number-depleting SIMP annihilation, is also
  possible~\cite{Bernal:2015ova}. Nevertheless, it only works for a
  very narrow parameter region, which will not be investigated here.}
Under the assumption that the dark sector is never thermalized, the
value of $R_0$ can be close to unity, provided the longevity of
state~2.  However, non-thermalization within the dark sector typically
requires small values of the dark coupling
$g_V\lesssim 10^{-3.5}\,\sqrt{m/\text{MeV}}\,(m_V/m)^{1/4}$  as obtained
from considering the rate for $VV\leftrightarrow 11\,(22)$.
In this case, the corresponding condition to achieve
$R_0 = O(1)$ is the same as Eq.~\eqref{relaFO:R}, but the temperature
ratio between two sectors now relies on the details of
freeze-in. Roughly, the upper limit on $g_V$ for freeze-in can be
approximated by setting $T'_\text{FO}\sim T_\text{FO}\sim m$ in
Eq.~\eqref{relaFO:R}, suggesting quite small dark couplings
$g_V \lesssim 10^{-3.5}\,m_V/(m\,\Delta m\,\text{MeV}^2)^{1/4}$.
Therefore, for both 4-to-2 SIMP annihilation and freeze-in,
$R_0\simeq 1$ cannot be achieved for sizable large dark couplings,
unless new ingredients such as dark radiation or non-standard
cosmology are considered.

\section{Astrophysical Implications of Split SIMPs}
\label{sec:astro}

So far we have studied how the DM relic abundance can be generated in
the early Universe for split SIMPs and computed the relative abundance
of the two states.  The assumed small mass splitting and the feeble
coupling to the SM sector suppress the decay width of the heavier
state, making a two-component DM model viable. In this section we
investigate the ensuing astrophysical implications.

\begin{figure}[h!]
\centering
\includegraphics[width=0.49\textwidth]{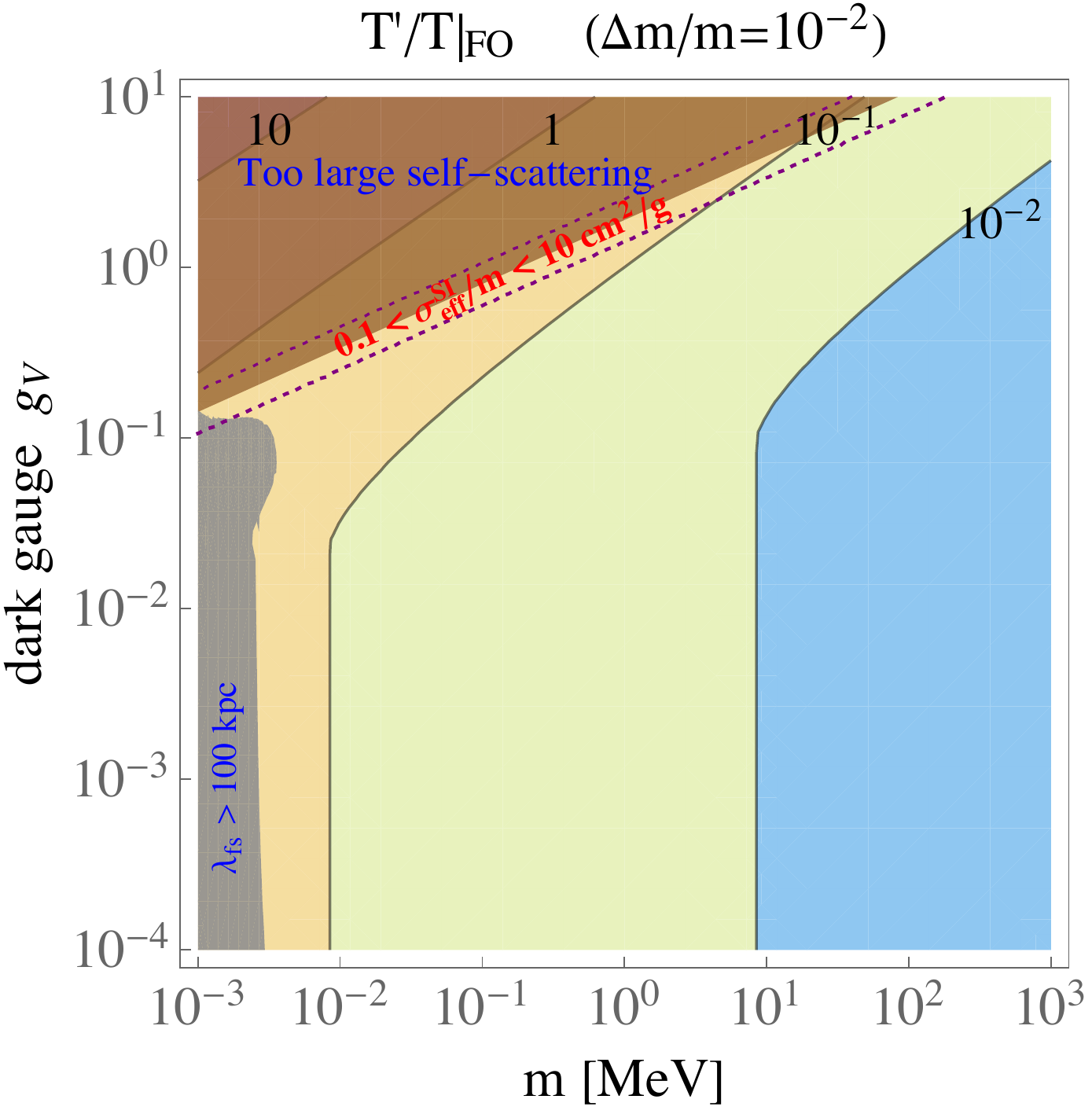}
\includegraphics[width=0.49\textwidth]{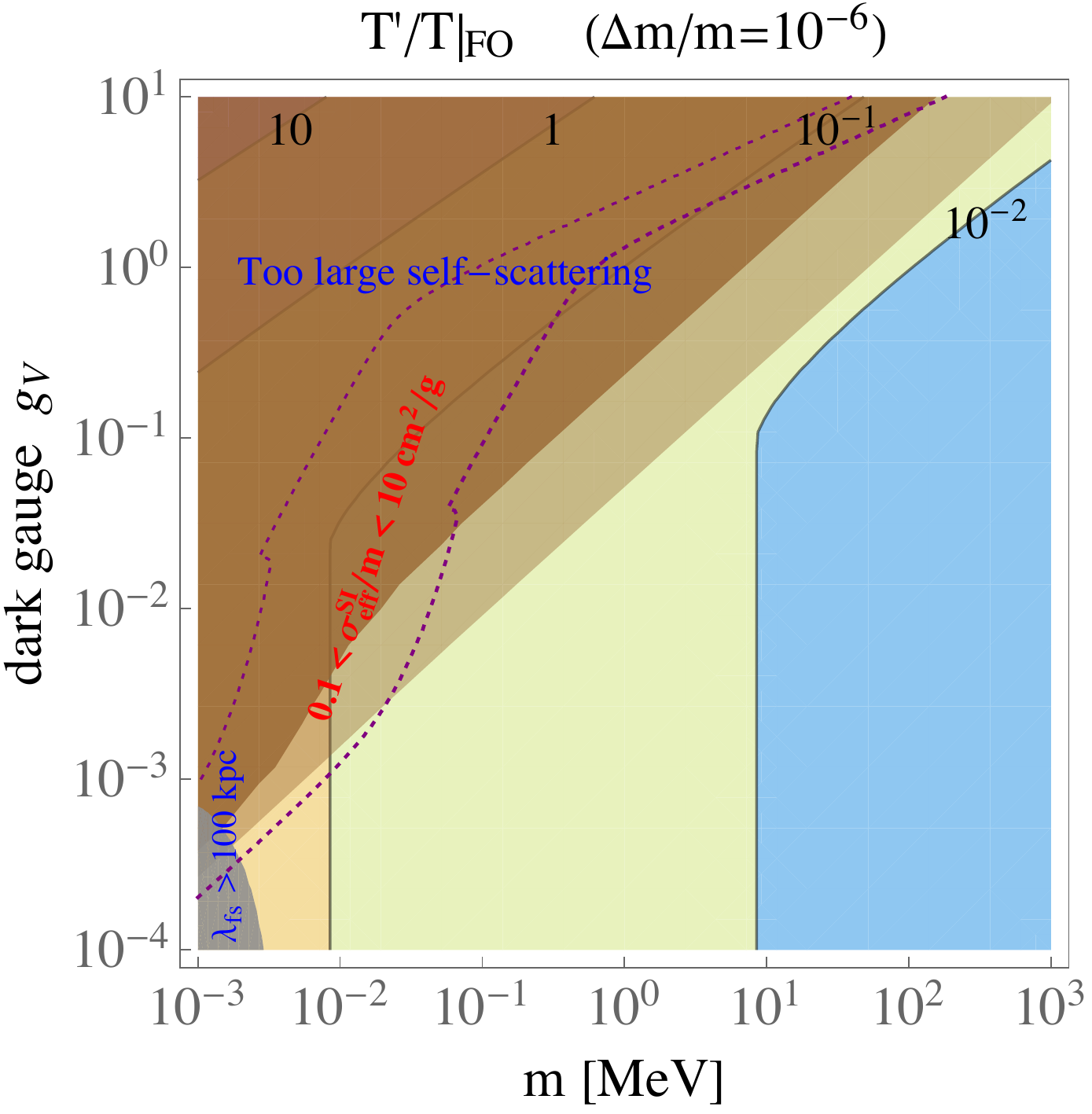}
\caption{Pseudo-Dirac DM assuming $\mzp=\frac52 m$ and
  $\Delta m/m =10^{-2}$\,(left) and $10^{-6}$\,(right). Contours show
  the temperature ratio of dark sector to visible sector at 4-to-2
  freeze-out; current constraints are shown by shaded regions, labeled
  in blue. In the dark (light) brown shaded region the
  \emph{effective} self-scattering cross section per DM mass is larger
  than 1~cm$^2$/g assuming $v_0 = 250$~km/s ($v_0 =1000 $~km/s). In
  addition, the window between
  $\sigma_\text{eff}^\text{SI}/m=0.1$~cm$^2$/g and $10$~cm$^2$/g for
  dwarf galaxies ($v_0=30$~km/s) is bounded by the dotted
  lines. Inside this region, small-scale structure problems can be
  addressed by DM.}
\label{fermionDM:SIbound}
\end{figure}

Due to the non-diagonal coupling in the dark sector, self-scattering
$11\to11$ only appears at the loop-level. Hence, the relative
abundance of states 1 and 2, $R_0$, enters in the DM self-scattering
$12\to12$ in dark halos of dwarf galaxies and colliding clusters as
well as in the determination of the free-streaming length of keV DM
affecting structure formation.
Even if state 2 decays, it may affect structure formation at small
scales due to the small velocity kick that the daughter DM particle
receives. %
These three effects are investigated separately below, and the results
are summarized in Fig.~\ref{fermionDM:SIbound} for
$\Delta m /m = 10^{-2}$ (left panel) and $10^{-6}$ (right panel).  While they are calculated for
pseudo-Dirac DM, similar results apply to scalar DM.

\subsection{Decaying DM after Decoupling}
\label{sec:decaying-dm-after}

The ratio of states 2 and 1 today, $R_0$, is obtained from the
value of decoupling by
$R_0 = R_{\rm dec}\, e^{-t_0/\tau_2} \sim R_{\rm dec}\, e^{-\Gamma_2/H_0}
$
where $\tau_2 = \Gamma_2^{-1}$ is the lifetime of state~2.  In the
minimal set-up, decays into a single on-shell photon final state,
$\phi_2 \to \phi_1 \gamma$ and $\chi_2 \to \chi_1 \gamma$, cannot
occur through kinetic mixing.%
\footnote{This can be seen from partial integration of
  $V_{\mu\nu}F_Y^{\mu\nu} = -2\,V_{\nu}\partial_\mu F_Y^{\mu\nu} =0$
  setting the photon on-shell.  Anomalous operators ($VFF,\, VVF$) are
  also negligible since the dark sector is non-chiral up to the mixing
  with the $Z$ boson. } 
State~2 hence decays to state~1 plus a pair of SM fermions, or three
photons, depending on kinematics.
The decay into an electron/positron pair $\chi_2 \to \chi_1\,V^{*} \to \chi_1\,e^+ e^-$ is allowed for
$\Delta m > 2\,m_e$ and typically dominates the total decay rate.
For kinematically
  unsuppressed decay,
\begin{equation}
  \Gamma_{\xt\to\xo e^+e^-} \simeq  \frac{2 \alpha\, \alpha_V\, \kappa^2}{15 \pi} \frac{ \Delta m^5}{ \mzp^4} 
  \simeq  2\,H_0 \times  \frac{m}{100~\MeV} \frac{\alpha_V}{\alpha} \left( \frac{\kappa}{10^{-10}} \right)^2 
  \left( \frac{\Delta m/m}{10^{-3}} \right)^5  \left( \frac{m}{m_V}   \right)^4 ,
\end{equation}
where $H_0$ is the Hubble constant, implying a lifetime that can be
comparable to the age of the Universe.  For scalar DM an extra factor of $2$ needs to be added. The total decay width to SM
neutrinos, $\Gamma_{\xt\to\xo\,\nu\bar\nu}$, is suppressed by another
factor of $(\mzp/m_Z)^4\times 3/(8\cos^4\theta_W)$.%
\footnote{A faster rate into neutrinos is possible if, in addition to
  kinetic mixing, $V$ and $Z$ mix through their mass terms,
  $\Gamma_{V\to \nu\bar\nu} \propto \delta_Z^2 m_V^3/m_Z^2$ where
  $\delta_Z \ll 1$ is a dimensionless number parametrizig the
  mixing~\cite{Davoudiasl:2012ag}.}
If the decay to charged lepton pairs is kinematically forbidden, the
decay proceeds dominantly via the emission of three photons,
$\chi_2 \to \chi_1\,V^{*} \to \chi_1\,3\gamma$,
\begin{eqnarray}
\Gamma_{\xt\to\xo3\gamma} 
& \simeq&  \Gamma_{\xt\to\xo \nu \bar \nu} \times  \left. \frac{ \Gamma_{\zp\to 3\gamma}}{\Gamma_{\zp\to \nu \bar \nu}}\right|_{\mzp \to \Delta m}   \notag\\
&\simeq &  H_0 \times  \left(\frac{m}{50\,\MeV} \right)^9 \frac{\alpha_V}{\alpha} \left( \frac{\kappa}{10^{-10}} \right)^2 
  \left( \frac{\Delta m/m}{10^{-2}} \right)^{13}  \left( \frac{m}{m_V}   \right)^4.
\end{eqnarray}
where we have made use of the expression for
$\Gamma_{\zp\to 3\gamma} $ as given in Ref.~\cite{Pospelov:2008jk} in the
second line. Owing to the strong dependence $\Delta m/m$ it shows that
for $\Delta m < 2\,m_e$ state 2 quickly becomes extremely
long-lived. Hence, it is always possible to realize
$R_0 \simeq R_{\rm dec}$ and we conclude that the model typically
features two-component DM where state~2 decays on cosmological time
scales.

Decaying DM has long been proposed as a solution to small scale
structure problems. For instance, the cosmological simulation in
Ref.~\cite{Wang:2014ina} has suggested that if the lifetime of a
heavier state is of the order of the age of the Universe, the kick
velocity $v_{\rm k}$ received by the daughter DM particle should be
around 20--40~km/s, or $(0.6-1.2)\times 10^{-4}c$, to appreciably
modify structure formation at comoving length scales of
${\mathcal O}$(100)~kpc.
The split SIMP provides a new example of this kind of models since
$v_{\rm k}/c \simeq \Delta m/m$ pointing to the $\Delta m/m =10^{-4}$
ballpark.
Finally, we note in passing that DM decay into radiation, without
seriously disrupting halo-structure (non-relativistic kick), is not
able to reduce the discrepancy between \emph{Planck} and low-redshift
measurements of $H_0$~\cite{Berezhiani:2015yta}. To address the last
discrepancy, both $R_0\sim 1$ and $\Delta m/m\sim 0.1$ are required.

\subsection{Self-scattering of SIMPs}

\begin{figure}[tb]
  \centering
\includegraphics[width=0.85\textwidth]{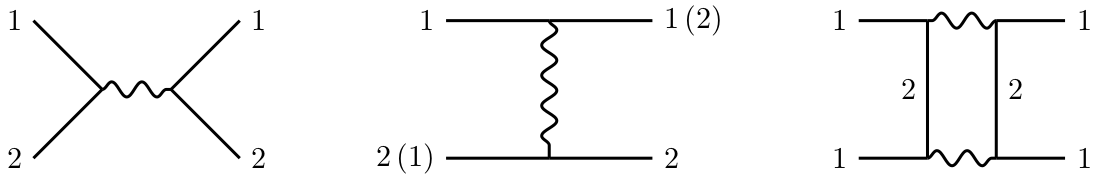}
\caption{\small Main processes contributing to DM self-scattering
  involving an initial state~1. The first two diagrams contribute to
  the scattering between two different states, $12\to 12$. The middle
  diagram can also contribute to inelastic up-scattering, $11\to 22$ and
  the last diagram shows the radiatively induced scattering
  $11\to 11$.}
\label{fig:selfscattering}
\end{figure}

After having established the conditions for longevity of the excited
state above, we now turn to its implications in the low-redshift
Universe. Figure~\ref{fig:selfscattering} shows the contributions to
the self-scattering involving an initial state~1.%
\footnote{The rate for self-scattering of states~2, $22\to11$ or
  $22\to22$, is suppressed by $R_0^2$ and correspondingly less
  important.}
They are $12\to12$ elastic scatterings, $11\to22$ endothermal
scatterings and radiatively induced $11\to11$ scatterings %
denoted as $\sigma_{12}$, $\sigma_\text{en}$ and
$\sigma_\text{rad}$, respectively.%
\footnote{The case of self-interacting inelastic DM where self-scatterings take place through light mediators has been studied in Ref.~\cite{Blennow:2016gde}.}
The cross sections for these
processes are listed in Appendix~\ref{app:sigmas} for both fermionic
and bosonic DM.

As explained in Sec.~\ref{sec:introduction}, a SIMP model that is able
to address small-scale structure problems should have a
self-interaction cross section of $\sigma/m\ge 0.1$ cm$^2/$g inside
dwarf galaxies.  In contrast, at larger scales astrophysical
observations suggest an upper bound on DM self-interaction.  Given the
uncertainties existing in the relevant
constraints~\cite{Agrawal:2016quu, Robertson:2016qef}, as a simple
criterion, we required the effective self-scattering cross section
involving state~1 over DM mass, as defined below, to be smaller than
1~cm$^2$/g at the galactic scale,
\begin{equation}
\label{eq:selfconstraint}
\frac{\sigma^{\rm SI}_{\rm eff}}{m} \equiv R_0 \frac{\sigma_{12}}{m} + \frac{{\langle \sigma_\text{en} v \rangle}  }{2\,m\, v}  
+ \frac{\sigma_\text{rad}}{2\,m} \lesssim {1~\cm^2/\rm g}\,,
\end{equation}
where a statistical factor of $1/2$ has been added to avoid
double-counting in the latter two terms. The first term includes a
factor of $R_0$ as the interaction rate for $12\to12$ hinges on the
abundance of states~2,
$ \Gamma_{12} \simeq R_0\,n_1\,\langle \sigma_{ 12} v \rangle\,.$
If state~2 has decayed, the remaining tree-level contribution to
self-scattering is given by the second term, describing the
endothermic scattering $11\to 22$.  Here, the relative kinetic energy
of initial particles with relative velocity $v$ must bridge the 
mass splitting,
\begin{equation}
{\langle \sigma_\text{en} v \rangle} \equiv \sigma^0_\text{en} 
 \iint  d^3\vec v_{1,2}\,f_1^\text{\tiny MB}\,f_2^\text{\tiny MB} 
 \sqrt{1 - \tfrac{2 \Delta m}{m v^2}  } |v |\simeq  \sqrt{\frac{2}{\pi}}
 \sigma^0_\text{en}v_{0}   \left[\xi\,e^{-\xi}\,K_1(\xi) \Theta\left(v_{\rm esc} - v_{0} \sqrt{\xi}\right)\right]\,,
\end{equation}
where $v_{\rm esc} $ is the escape velocity and $v_0$ is related to
the velocity dispersion of the system, $v_0 = \sqrt{2} \sigma$;
$f_i^{\rm MB} = N e^{-\vec v_i^2/v_0^2}$, normalized to unity. The Heaviside step
function $\Theta$ provides the kinetic cutoff for up-scattering. Here
$K_1(\xi)$ is the Bessel function of the second kind, with
$\xi\equiv 2\,\Delta m/(m v_0^2)$; hence, the last factor in the
square bracket converges to $1$ for $\xi \to 0$, and becomes
exponentially suppressed for $ \xi \gg 1$.
The third term in Eq.~\eqref{eq:selfconstraint} accounts for the radiatively induced elastic scattering
$11\to11$.  It is dominated by diagrams with the exchange of two
(heavy) $V$ bosons, and has an interaction rate
$\Gamma_{\rm rad} =n_1 \langle \sigma_{\rm rad} v \rangle $.  This process can
be important for certain parameter regions where a large mass
splitting significantly reduces the contributions of the two previous
contributions.

In Fig.~\ref{R0:msplit}, we identify the region in the $[g_V,\,m]$
plane which yields $R_{0}\,\sigma_{12}/m=1$~cm$^2$/g by the dashed
lines at the top-left corners. The left (right) panel corresponds to
$\Delta m/m= 10^{-2}$ $(10^{-6})$.  The value of $R_0$ is computed from
the decoupling and $R_0 = R_{\rm dec}$ has been assumed.  As a
comparison, parameters associated with $\sigma_{12}/m=1$~cm$^2$/g are
also given by the dotted line in each panel, highlighting the
importance of tracking $R$ until the decoupling of the 2-to-2 process
in the early Universe.
The potentially excluded regions from self-scattering,
  $\sigma_\text{eff}^\text{SI}/m>1$~cm$^2$/g at the galactic scale (at
  the cluster scale) are given by the dark (light) brown shaded regions
  in both panels of Fig.~\ref{fermionDM:SIbound}. For the left panel,
  both exclusions coincide because the endothermic channel is always
  strongly suppressed due to the large mass splitting. Meanwhile,
$\sigma_\text{eff}^\text{SI}/m=0.1$~cm$^2$/g and $10$~cm$^2$/g for
dwarf galaxies are shown with two dotted lines; the region in between
addresses the aforementioned small-scale structure problems.

It is worth emphasizing that the constraint~\eqref{eq:selfconstraint}
has to be viewed with some caution.  The abundance ratio of the two
species is assumed to be universal, and is set to be $R_0$.  In
practice, this ratio may depend on the dynamics and merger history of
dark halos.  On the one hand, dense DM halos with large velocity
dispersion may enhance the $11\to 22$ endothermic process, thereby
increasing $R_0$ differentially in those objects.  On the other hand,
as state~2 only contributes a small portion of the total DM abundance,
scattering with state~1, during halo mergers, may deplete state~2 in
the inner regions of halos, hence suppressing the overall effect of DM
self-interactions.  Accounting for the latter likely requires
cosmological simulation and is beyond the scope of this
work.

\subsection{Free Streaming}
\label{freestreaming}

Self-scattering is also relevant in the high-redshift Universe as it
controls the free-streaming of DM particles, which in turn determines
the smallest DM objects that can be formed from primordial
perturbations~\cite{Profumo:2006bv}. Here, two competing effects need
to be considered. First, in the 4-to-2 annihilation process, DM
particles get boosted in the final state. The dark sector hence has a
tendency to be naturally hotter than the SM after freeze-out, adding
to the free-streaming after DM chemical decoupling. Second, 2-to-2
self-scattering will keep the particles contained up to diffusive
processes until a temperature $T'_k$ when the latter cease,
\begin{equation}\label{kindec} {\Gamma_1(\tpk)} \sim {H(\tk)}
\end{equation}
where $\Gamma_1=  n_1 \langle  \sigma^{\rm SI}_{\rm eff} v \rangle$ 
for state~1. Apparently, this happens after the decoupling of the
annihilation process $22\to
11$, that is, $T_{\rm k}
<T_\text{dec}$.  Since the interactions are short-range, $m_V >
m_1+m_2$, we are allowed to neglect the detailed momentum transfer in
each collision~\cite{Bringmann:2006mu}.

After the DM kinetic decoupling at cosmic time $t_{\rm k}$, the
(non-relativistic) DM velocity redshifts as
$v_\chi(T') = (T/\tk)\, v_{\chi}(\tpk) $, where
$v_{\chi}(\tpk)\simeq \sqrt{{2\,\tpk}/{\mx}}$. The comoving
free-streaming length $\lfs$ is given by integrating the velocity
until matter-radiation equality at cosmic time~$t_{\rm eq}$,
\begin{equation}
\lfs= \int_{t_\text{k}}^{t_\text{eq}}\frac{v_\chi(t)}{a(t)}dt
\sim \frac{26\,\text{kpc} }{\sqrt{\gs(T_k)}} \times \frac{10\ \keV}{\sqrt{T_k\, m}}   \left( \frac{\tpk }{\tk}  \right)^{1/2} \log_{10}\left( \frac{T_k}{T_\text{eq}} \right) \,.
\label{fsl}
\end{equation}
Here, we assumed that the Hubble rate $H$ is dominated by the SM
energy density and $\gs(T)$ counts the SM relativistic degrees of
freedom at photon temperature $T$; $T_0$ and $\teq$ are the
corresponding temperatures at present and at matter-radiation
equality, respectively.  In our numerical evaluation, we additionally
take into account the free-streaming in the matter-dominated epoch.

The strongest observationally inferred limit on $ \lfs$ is derived
from the matter power spectrum suppression induced by DM
free-streaming and comparing Lyman-$\alpha$ observations and
cosmological hydrodynamical simulations. The current limit is
$\lfs\lesssim 100$~kpc~\cite{Viel:2013apy,Baur:2015jsy} and it is
shown by the gray shaded region in Fig.~\ref{fermionDM:SIbound}.  As
can be seen, the bound for sizable $g_V$ is very weak due to the
efficient DM self-scattering.
Finally, we point out that there are additional effects from
collisional damping and dark acoustic
oscillations~\cite{AtrioBarandela:1996ur}, which are not considered
here because they only lead to a suppression of the matter power
spectrum at similar or scales smaller than $\lfs$ for our parameter
choices. They become important in the limit $m_V \ll m$ for which
additional dark radiation in form of $V$ remains coupled with
DM~\cite{Cyr-Racine:2013fsa}.
\section{Sensitivity and Constraints on the Kinetic Mixing Portal}
\label{sec:connection}

\begin{figure}[t]
\centering
\includegraphics[width=0.55\textwidth]{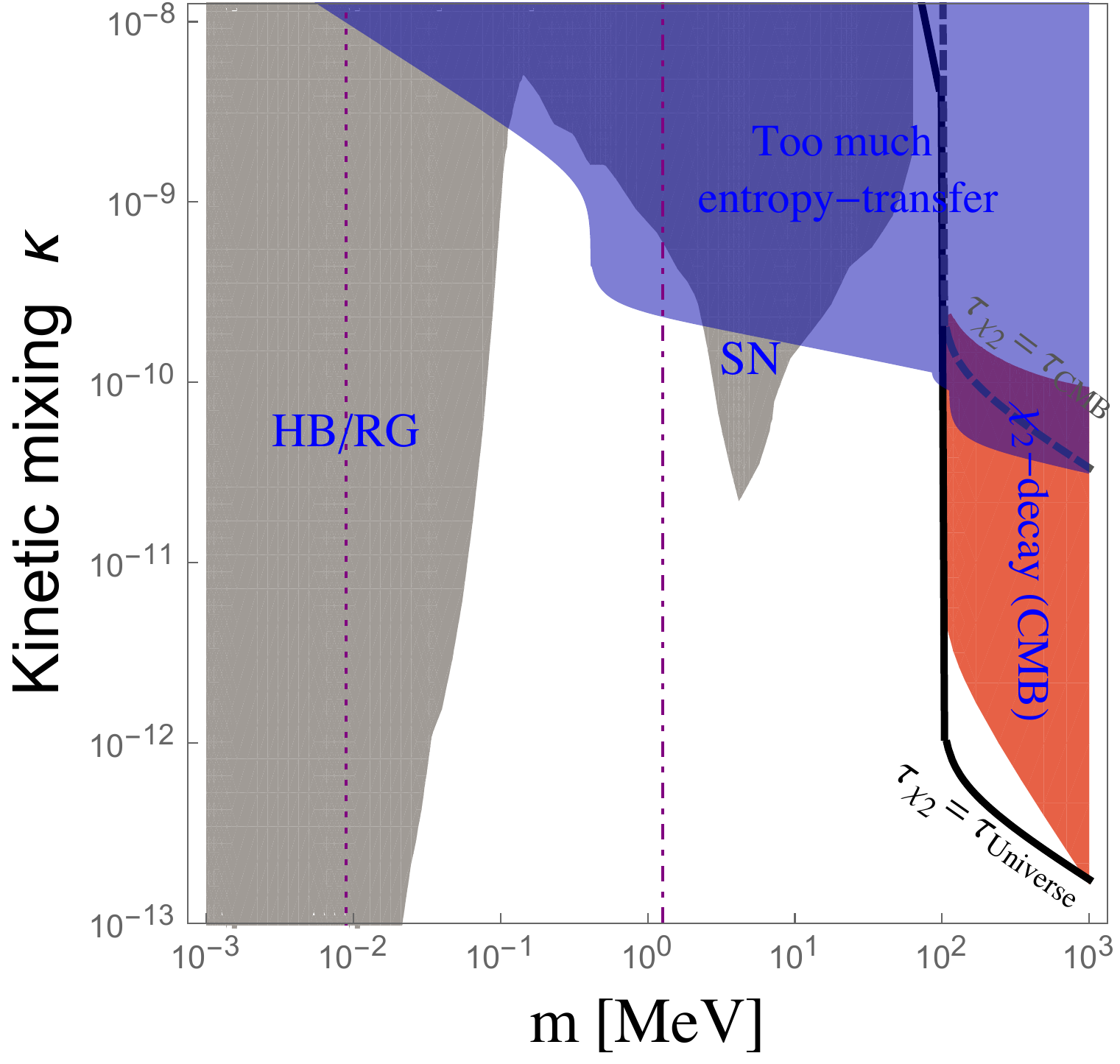}
\caption{Bounds on the kinetic mixing parameter $\kappa$ for
  $\Delta m/m = 10^{-2}$ and $g_V = e$ for pseudo-Dirac DM. The mass
  of $V$ is set to be $\frac52 m$. To the right of the solid (dashed)
  thick contour lines the lifetime of state~2 falls below the age of
  the Universe (the cosmic time at last scattering). All shaded
  regions with labels in blue are excluded (see text for details.)
  The two thin vertical lines show DM
  self-interaction bounds for $\Delta m/m = 10^{-2}$ (dotted) and
  $10^{-6}$ (dash-dotted), respectively. The corresponding regions to
  the left are disfavored.
}
\label{fig:kmbounds}
\end{figure}

By assumption, the dark sector interacts with the SM through the
``kinetic mixing'' portal. The requirement on maintaining decoupled
sectors in the early Universe, and on ensuring longevity of state~2,
implies $\kappa \ll 10^{-6}$, precluding missing energy searches in
both collider and fixed-target experiments. Although astrophysics
remains the best probe, indirect and direct searches have some
potential to probe the model by exploiting the mass splitting between
the two states. Below we provide a brief discussion thereof, and for
exemplary mass-splitting $\Delta m/m = 10^{-2}$ and $g_V=e$, the
parameter space in $[m,\,\kappa]$ is chartered out in
Fig.~\ref{fig:kmbounds}.

\begin{itemize}

\item Equation~\eqref{entropy:ratio} relates the relative comoving
  entropies of the hidden and SM sector, $S'/S \ll 1$, through the DM
  relic abundance requirement. A priori, this ratio can be originated from
  sources independent of the SM, such as a branching of the inflaton decaying into
  the dark sector. Any finite value of $\kappa $ will lead to the
  transfer of energy/entropy from the SM to the dark sector, hence
  increasing the ratio $S'/S$ and thereby imposing an upper limit on
  the strength of the portal interaction.
  Notice that the energy transfer is significantly enhanced if $V$
  bosons can be created on-shell by electron pair annihilation,
  \textit{i.e.}, $m_V \ge 2\,m_e$. This implies
  $\kappa \lesssim {\mathcal O}(10^{-10})$ for MeV DM with a heavier
  $m_V$ and no additional contributions to the DM abundance.
  In turn, for $m_V < 2\,m_e$ the value of $\kappa$ is allowed to
  increase into observationally more interesting regimes. The excluded
  region assuming no other a priori sources than SM for dark sector
  particles is labeled ``too much entropy-transfer'' in
  Fig.~\ref{fig:kmbounds}.

\item The decay of state~2 into state~1 is accompanied by SM
  radiation, injecting an energy per baryon of
  $ E_{\rm inj}/n_B = \Delta m\, (R_{\rm dec}\, n_1)/n_B $, possibly constrained by
  BBN and CMB. For BBN, the ballpark of sensitivity on electromagnetic
  energy injection is $O(\MeV)$/baryon for $\tau_2\gtrsim 10^5\,\sec$,
  see \textit{e.g.} Ref.~\cite{Pospelov:2010hj}.  Hence, only
  mass-splittings $\Delta m \gtrsim \MeV > 2\,m_e$ are energetically
  relevant, where the decay proceeds by producing an electron-positron
  pair. We conclude that, approximately,
  \begin{align}
    \frac{\Delta m}{m} \lesssim \frac{10^{-3}}{5 R_{\rm dec}\, {\rm Br}_{\rm em}  \,e^{-t/\tau_2}}
  \end{align}
  should hold for respecting the limits on electromagnetic energy injection (with
  branching ${\rm Br}_{\rm em}$) from BBN; the factor $5$ is the ratio
  $\Omega_{\rm DM}/\Omega_b$ of DM to baryon densities today,
  assuming that decoupling in the dark sector happens before
  $t\sim 10^5\,\sec$. It turns out that in our model the BBN
  constraints are respected.
  Longer lifetimes are most stringently constrained by CMB
  measurements~\cite{Ade:2015xua, Slatyer:2015kla, Slatyer:2016qyl},
  rather than by indirect searches~\cite{Essig:2013goa}. The limits
  derived in Ref.~\cite{Slatyer:2016qyl} can be cast into
  \begin{align}
   \frac{\Delta m}{m} \lesssim \frac{\tau_2}{10^{24}\,\sec} \frac{1}{{\rm Br}_{\rm em}  \,R_{\rm dec}}\,,
  \end{align}
  which is shown as the orange shaded region on the right-hand side of  Fig.~\ref{fig:kmbounds}. As can be seen from the figure, when the
  channel to electrons becomes kinetically forbidden at $m \lesssim 100$\,MeV, $\tau_2$
  increases dramatically and surpasses this bound.

\item The heavier state behaves as exothermic DM and may thus be
  detectable in direct detection experiments. Although its mass lies
   below the GeV scale, the energy transfer to the target~A
  (nucleus or electron) through the inelastic collision
  $2+ A \to 1 + A $,
\begin{equation}
E_\text{recoil} \sim \Delta m \, \frac{\mu_{mA}}{m_A}\,, %
\end{equation} 
where $\mu_{mA}$ is the reduced mass between DM and target, can be
larger than the experimental threshold. The scattering results in a
monochromatic signal, up to corrections proportional to the relative
kinetic energy and bound state effects (the latter of which can be
substantial). For the small mass splittings of interest, collisions with
electrons are more prospective, as they cause larger energy
transfer. The cross section on the latter is given by
\begin{align}
  \bar \sigma_e \simeq a\, 10^{-44} \, \cm^2 \times
  \frac{\alpha_V}{\alpha} \left( \frac{\kappa}{10^{-10}} \right)^2 
 \left(  \frac{m}{100\,\keV}\right)^2  \left(\frac{300 \,\keV}{m_V}  \right)^4 \,,  
\end{align}
where $a = 1$ $(1/2)$ for fermions (scalars) and assuming $m\lesssim m_e$.
Since we consider $m_V > 2\, m$, the cross section is suppressed and
current sensitivity from ionization only data from
XENON10~\cite{Angle:2011th} and XENON100~\cite{Aprile:2016wwo} falls
short; DM limits based on the latter data for elastic scattering and
absorption have previously been presented
in Refs.~\cite{Essig:2012yx,An:2014twa,Bloch:2016sjj,Kouvaris:2016afs}. However,
once $\Delta m \gtrsim 1\,\keV$ the energy deposition is large enough
for reducing the background via the scintillation signal. Adopting the
results from an axion search in XENON100~\cite{Aprile:2014eoa} we find
that direct detection starts probing the plane for
$\sqrt{R_{0}}\,\kappa \gtrsim 10^{-9}$ %
with improvements expected from
LUX~\cite{Akerib:2015rjg} and XENON1T~\cite{Aprile:2015uzo}. Finally,
a whole number of proposed searches may be able to improve the
sensitivity on $\kappa$---and to be sensitive to minute
$\Delta m \lesssim \eV$---in the
future~\cite{Graham:2012su,Essig:2015cda,Guo:2013dt,Schutz:2016tid,
  Hochberg:2015pha, Hochberg:2015fth}; see
Appendix~\ref{sec:exoth-dm-electr} for details on the derivation and more results of
the direct detection region.

\item Finally, astrophysical constraints are of significant relevance
  at probing the production and emission of feebly interacting light
  particles from stars, such as horizontal branch (HB) and red giants (RG), as well as supernovae (SN)~\cite{Raffelt:1996wa}. For
  $m\lesssim 100\, \keV$, energy losses from stars by on-shell production
  of $V$ dominate the constraints on $\kappa$,
  see Refs.~\cite{An:2013yfc,Redondo:2013lna,An:2014twa}. MeV-scale DM is
  constrained by SN where we adopt the latest bound from
  Refs.~\cite{Chang:2016ntp, Hardy:2016kme}. The astrophysically
  disfavored regions are shown by the gray shaded regions as labeled.
  Note that in our scenario $V$ decays promptly and dominantly into
  lighter dark particles in contrast to (meta-)stable $V$-bosons
  considered in Refs.~\cite{Essig:2013goa, Fradette:2014sza}.

\end{itemize}

Most of the parameter space shown in Fig.~\ref{fig:kmbounds}
corresponds to a state 2 with a lifetime that exceeds the age of
the Universe.  Only in the small region delimited by the solid black
line in the top-right corner, the heavy state effectively decays.  As
discussed in the previous section, the abundance of state~2 plays an
important role in addressing small-scale problems if $\sigma_{12}/m$
is large. 
All shown constraints can be easily rescaled to values of $g_V$ from
$\sim 10^{-4}$ to $\mathcal{O}(1)$.  The bounds from stars (HB and RG)
as well as from SN (gray shaded region) are
independent of $g_V$; for as as long as $V$ decays dominantly within
the dark sector, the lifetime of $V$ does not enter the energy loss
argument. The constraint on entropy-transfer (blue shaded region) is
derived under the conservative condition that the dark sector becomes
populated only through the portal interaction.  The value of $g_V$ is
only relevant when $m_V \lesssim 1$\,MeV and $g_V \gtrsim e$, as
otherwise the on-shell production of vectors $V$ dominates the
entropy-transfer. For bounds from indirect (and future direct)
searches, one has to take into account the corresponding change of
$R_0$; smaller values weaken the constraints on $\kappa$. Since
$R_0 \propto g_V^{-4}$, obtained from Eq.~\eqref{eq:de}, the y-axis can
be replaced by $(e/g_V)\,\kappa $.
Finally, the limits on self-scattering are not easily rescaled with
$g_V$ but are shown as two vertical purple lines for
$\Delta m/m = 10^{-2}$ (dotted) and $10^{-6}$ (dash-dotted). For each
mass splitting value and $g_V = e$, parameter region on the left side
of the line has been excluded. Such limit on $m$ weakens with the decrease of $g_V$.

\section{Conclusions and Outlook}
\label{sec:conclusions}

In this paper we have considered a variant of strongly (yet
perturbatively) self-interacting DM models in which scattering rates
are controlled by a fine mass splitting $\Delta m/m $ between an
otherwise degenerate pair of real scalars $\phi_{1,\,2}$ or Majorana
fermions $\chi_{1,\,2}$. The gauged dark U(1) symmetry with vector $V$
and gauge coupling $g_V$ is explicitly broken by $\Delta m$ and thus the
interaction with DM proceeds off-diagonally.

We first calculate the DM relic abundance in 4-to-2 number-depleting
processes. It is well known that the requirement for successful 3-to-2
or 4-to-2 freeze-out points towards sub-GeV DM, and we study the
principal DM mass range from keV to GeV. In order to avoid constraints
on dark radiation during BBN, we consider a scenario in which the dark
and observable sector do not reach thermal equilibrium.
Non-relativistic four-fermion annihilation is inhibited by the
Pauli exclusion principle and requires two spin-singlet pairs of
$2\chi_1$ and $2\chi_2$. The efficiency of annihilation is hence regulated
by the number ratio $R=n_2/n_1 \sim e^{-\Delta m/T'}$, which, in turn
puts a requirement on the minimum value of $g_V$ for given $\Delta m$
and given ratio of photon and dark sector temperatures at freeze-out
$T'/T|_{\rm FO}$. In contrast, for scalars, we find the required value
of $g_V$ to be largely independent of $\Delta m$ as $4\phi_1$
annihilation channels are open.

We then track the ratio $R$ until it is frozen out in the decoupling
of the $12\leftrightarrow 12$ process. This sets the boundary
condition for the self-scattering in DM halos today: whereas the
$11\to 22\, (11)$ scattering is kinematically (radiatively)
suppressed, the scattering $12\to 12$ is regulated by the value of
$ R_0$ in the low redshift Universe. The latter is determined by the
lifetime of state~2, $\tau_2$, through
$R_0 = R_{\rm dec} e^{-t_0/\tau_2}$. We choose a kinematic setting
$m_V \ge 2m +\Delta m$ so that $V$ decays promptly within the dark
sector; $\tau_2$ is then determined by the off-shell decay
$2\to 1 + V^{*} \to 1 + \rm SM$. For this we couple $V$ to SM through
kinetic mixing with strength $\kappa$, and for $\Delta m \leq 2m_e$
state~2 becomes quickly stable  on cosmological timescales.

Since freeze-out in the dark sector through number-violating processes
can lead to hot DM, it is usually surmised that the coupling to SM must
bring DM into kinetic equilibrium with SM, putting a lower bound on
$\kappa$. Here, we study an alternative scenario in which the dark
sector remains decoupled from SM, but free-streaming constraints are
respected 1) by the containment of particles through self-scattering
(with diffusion playing a sub-leading role) and 2) by a freedom of
choice in the initial temperature ratio $T'/T$ of both sectors.

Astrophysical bounds on the scenario are controlled by $\kappa$ as it
regulates the emission of light DM states through the production of
$V$ and the allowed parameter space is identical to the one for
long-lived ``dark photons''.  We identify the parameter regions where
the relic density requirement is matched, and where, at the same time,
a self-scattering cross section of $0.1\lesssim\sigma/m\lesssim10$
cm$^2/$g is attained. The latter ballpark is proposed to solve the
small-scale problems in the collisionless DM paradigm. We find that it is
possible to reach such regions in parameter space while respecting all
astrophysical and cosmological constraints. %

In a final step, we study the detectability of the scenario in
terrestrial experiments. The requirement that dark and observable
sectors remain decoupled in the early Universe, puts collider probes
out of reach. However, the exothermic $2\to 1$ scattering on
electrons leads to a quasi-monochromatic energy deposition, enhancing
the prospects in direct detection experiments. We find that provided that 
$\Delta m > 1\,\keV$ and $R_0\sim 1$, the scenario is on the verge of being probable
with reported data (S1 and S2) from XENON100, despite the requirement
$\kappa \lesssim 10^{-9}$. Ionization only (S2) studies are not
sensitive at this point.

Several points raised in the paper prompt further
investigation. First, we specialized to the case $m_V > 2m +\Delta m$,
which is a restrictive requirement for both efficient annihilation,
and detectability in underground rare event searches. It was largely
motivated by evading constraints on extra radiation and from
astrophysics, however, a more detailed exploration of the parameter
space is certainly warranted, particularly in light of upcoming CMB
measurements with much improved error bar on dark radiation,
$\sigma(N_{\rm eff}) =
O(10^{-2})$~\cite{Abazajian:2016yjj,DiValentino:2016foa}.
Second, as we have shown, future direct detection experiments will be 
sensitive to this physics, despite the requirement that dark and
observable sector never reach thermal equilibrium and the above limit
of a ``heavy'' mediator. Following recent proposals to employ
semiconductor targets with sensitivity to $O(\eV)$ energy depositions,
among others, a more general study of exothermic DM scattering with
astrophysically motivated mass splittings
$\Delta m /m = O(v^2)\sim 10^{-6}$ as well as lighter mediators is
certainly motivated. We leave such and related points for future work.

\section*{Acknowledgments}
NB thanks F.~Staub and A.~Vicente for helpful discussions.  NB is
partially supported by the São Paulo Research Foundation (FAPESP)
under grants 2011/11973-4 \& 2013/01792-8, the Spanish MINECO under
Grant FPA2014-54459-P and by the ``Joint Excellence in Science and
Humanities'' (JESH) program of the Austrian Academy of Sciences.
XC and JP are supported by the
``New Frontiers Program'' by the Austrian Academy of Sciences. This
project has received funding from the European Union's Horizon 2020
research and innovation programme under the Marie Skłodowska-Curie
grant agreements 674896 and 690575. 

\appendix

\section{Annihilation Rates}
\label{sec:annihilation-rates}

Here we provide some details on the rates that go into the Boltzmann
equation for the 4-to-2 process. The squared matrix elements
$|\overline M|^2$ are obtained in the usual way and, for fermions, are
summed (averaged) over final (initial) state spins; a symmetry factor
of $1/2$ is included for identical final state particles.  For reasons of a
tractable exposition, we only list the leading order in relative
velocities and $\Delta m$, which means a constant in both
quantities. Therefore, the phase space integration is trivial, the
squared matrix elements only depend on the initial state
configuration, and the rates are directly obtained from the latter
through,
\begin{align}
\langle ijkl\to mn\rangle & =
  \frac{S_{ijkl}}{m_i m_j m_k m_l}\, \frac{1}{32\pi}\, \frac{p_f}{\sqrt{s}}\,|\overline M_{ijkl}|^2\, ,\\
  & \simeq \frac{\sqrt{3}\,S_{ijkl}}{128 \pi\, m^4}\,  |\overline M_{ijkl}|^2 \,.
\end{align}
where $p_f$ is the magnitude of the final state relative momentum and
$\sqrt{s}$ is the total incoming energy; $S_{ijkl}$ is a statistical
factor comprising a symmetry factor
($1/n!$ for each set of $n$ identical initial state particles) times the
multiplicity of particles that are removed from the plasma in the
annihilation (2 in this case).  Detailed balancing allows to make contact with an
actual 2-body cross section $ \langle \sigma_{mn}v\rangle $ of
particles $m$ and $n$ in the 2-to-4 process%
\begin{align}
\langle ijkl\to mn\rangle =2\, \langle \sigma_{mn}v\rangle \frac{n_m^\text{eq}n_n^\text{eq}}{ n_i^\text{eq} n_j^\text{eq}   n_k^\text{eq} n_l^\text{eq}} \,,
\end{align}
in which the prefactor of 2 gives the particle number changed by per reaction. In the case $m=n$, another factor $1/2$ needs to be added to the RHS to avoid over-counting. 

For pseudo-Dirac DM the only non-zero reactions for $ijkl\to mn$ are
$2\chi_i + 2\chi_j \to 2\chi_i$ with $i\neq j$,
\begin{equation}\label{eq:M2fer}
|\overline M_{1122}|^2 \simeq 216\,\gd^8\,\frac{\mx^4\,(\mzp^4-8\mx^2\mzp^2-8\mx^4)^2}{(\mzp^4-2\mx^2\mzp^2-8\mx^4)^4}\,;
\end{equation}
for scalars, we find
\begin{equation}\label{eq:M2sca}
|\overline M_{ijkl}|^2 \simeq c_{ijkl}\, \frac{ g_V^8 m^4}{(2m^2 + m_V^2)^4}\,.
\end{equation}
with $c_{1122} = 72 $, $c_{1111} = c_{2222} = 5184$, $c_{1112} =c_{1222} = 432$.  %
Equations~\eqref{eq:M2fer}
and~\eqref{eq:M2sca} are exact in the limit $\Delta m=0$, in the
numerical analysis we have used however the full expressions. The
sizable number of matrix elements were computed by implementing the
models into~\texttt{CalcHEP}~\cite{Belyaev:2012qa} and
\texttt{FeynArts}~\cite{Hahn:2000kx}.

\section{Solution to the Boltzmann Equation}\label{app:boltzmann}

At freeze-out, the DM density starts exceeding from its equilibrium
value, $Y\gg Y_\text{eq}$, so that the term proportional to
$Y_\text{eq}$ can be dropped in Eq.~\eqref{eq:Y} to obtain the yield
of DM at present,
\begin{equation} {Y_0}\simeq
  \left[\frac{1}{Y(\xfo)^3}+3\int_\xfo^\infty\frac{s^3\,\langle\sigma
      v^3\rangle_{4\to2}}{x\,H}dx\right]^{-1/3} \simeq
  \left[3\int_\xfo^\infty\frac{s^3\,\langle\sigma
      v^3\rangle_{4\to2}}{x\,H}dx\right]^{-1/3}\,,
\end{equation}
where  $Y_0\ll Y(\xfo)$ is also used in the last step.
The observed relic density parameter today~\cite{Ade:2015xua}
reads,%
\footnote{Assuming that the state 2 decays into 1 after the DM
  freeze out, with small correction proportional to $\Delta m/m$ when
  state~2 is stable.}
\begin{equation}
\Omega_{\rm DM} = \frac{m \,s_0\,Y_0}{\rho_c}=\left(2.742 \times 10^8 \,\text{GeV}^{-1} \, h^{-2}\right) \, m \, Y_0,
\end{equation}
where $s_0$ and $\rho_c $ denote the respective SM entropy density and critical energy
density today.
\section{Decay Rates and Self-scattering Cross Sections}\label{app:sigmas}

For DM self-scattering where state 2 has a lifetime much in excess of the age of the Universe,
DM may self-scatter entirely elastically, $12\to 12$, with cross
section: 
\begin{align}
\text{Pseudo-Dirac:}& \quad   \sigma_{ 12} =\frac{\gd^4 m^2}{4 \pi m_V^4} \left[ \frac{16 m^4 - 20 m^2 m_V^2 + 7m_V^4}{(4 m^2 - m_V^2)^2}
\right], \\
\text{Scalar:}& \quad   \sigma_{ 12} = \frac{\gd^4}{4\pi}\frac{m^2}{m_V^4}.
\end{align}

The cross section for the endothermic scattering 11$\to$22 is given by
(neglecting contributions proportional to $\delta^2$)
\begin{align}
\text{Pseudo-Dirac:}& \quad  \sigma_{en} = \frac{\gd^4 m^2}{8\pi m_V^4}  \sqrt{1 - \frac{2 \Delta m}{m v^2}  } \equiv  \sigma^0_{en}  \sqrt{1 - \frac{2 \Delta m}{m v^2}  }, \\
\text{Scalar:}& \quad   \sigma_{en} = 4 \sigma^0_{en}  \sqrt{1 - \frac{2 \Delta m}{m v^2}  } .%
\end{align}
The kinetic condition for the reaction is ${\Delta m} < {m v^2/2} $, where $v$ is the DM velocity in the center-of-mass frame.

Radiatively induced elastic scattering $11\to11$ can be important if
the above two processes are both suppressed, either kinematically
and/or by deficiency of reactant~2. From naive dimensional analysis
one may estimate it as,
\begin{equation}
\sigma_{\rm rad} \sim  \frac{\gd^8 m^6}{256\pi^4 m_V^8} .
\end{equation}
Finally, for completeness, we also list the decay width of the heavy
gauge boson,
\begin{eqnarray}
  \Gamma_{\zp\to\xo\xt}&=&
  \frac{\gd^2\mzp}{12\pi} \left( 1+\frac{\Delta m^2}{2m_V^2} \right)
\left[ 1 - \frac{(2m+\Delta m)^2}{m_V^2}   \right]^2 .
\end{eqnarray}

\section{Exothermic DM-Electron Scattering}
\label{sec:exoth-dm-electr}

\begin{figure}[t]
\centering
\includegraphics[height=7cm]{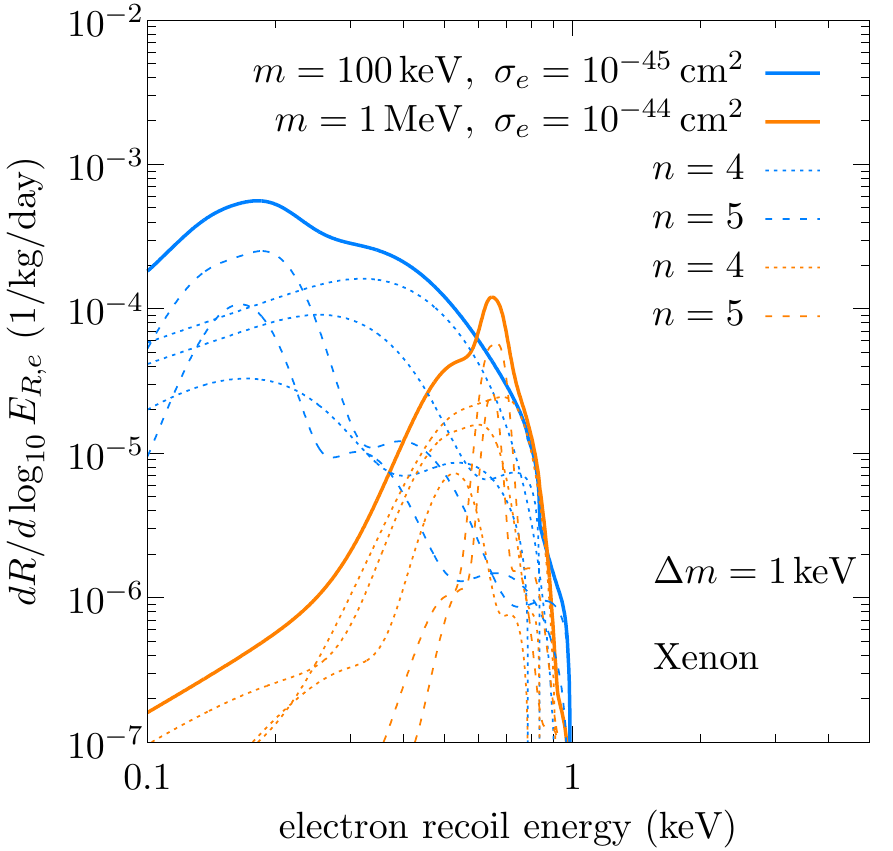}\hfill%
\includegraphics[height=7cm]{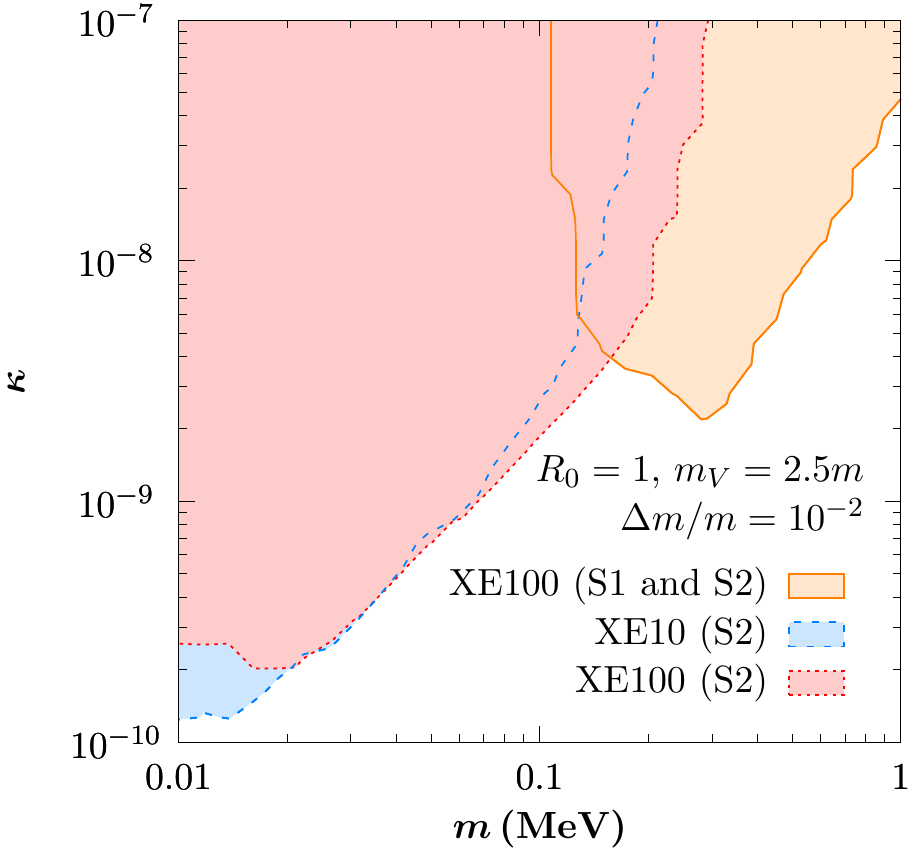}%
\caption{\textit{Left:} Exothermic scattering on electrons and
  resulting recoil spectra in a liquid xenon detector for a 1\,\keV\
  mass splitting and a combination of DM mass and scattering cross
  section $(100\,\keV, 10^{-45}\,\cm^2)$ and
  $(1\,\MeV, 10^{-44}\,\cm^2)$. \textit{Right:} The parameter space of
  $\kappa$ vs.~$m$ with current sensitivity in direct detection using
  the same choice of parameters as in Fig.~\ref{fig:kmbounds} but
  equipartitioning the abundance of state 2, $R_0 = 1$.}
\label{fig:dd}
\end{figure}

Here we provide the formul\ae\ for the exothermic scattering of state
2 to 1 on electrons, applicable to direct detection studies. A bound
electron has a fixed energy $E_e = m_e - E_B$ with $E_B > 0$ being the
magnitude of the binding energy, but a continuous distribution of
momenta $\vec p_e$.
The process we are considering is ionization, where the final state
electron has kinetic recoil energy $E_{R,\,e} $. When $\Delta m $ is
larger than the typical kinetic energy of DM, the signal will
correspond to a mono-chromatic energy deposition, with differential
cross section,
\begin{align}
\frac{ d \sigma v }{ d\ln E_{R,e}  }  &  =
  \bar \sigma_e  \frac{ m }{4\,\mu_{\chi e}^2  } 
\times  \bigg[ \int d\Omega_{\vec p_e'} \, q |F_{\rm DM}(q)|^2 
  \sum_{\rm deg.\,states} |f_{\rm ion}(q)|^2   \bigg]_{q = \sqrt{2 m\, ( \Delta m - E_B - E_{R,e} )}} 
\end{align} 
where $q= |\vec q|$ is the magnitude of the transferred three-momentum
$\vec q = \vec p_2 - \vec p_1$, \textit{i.e.},~the difference in DM
momenta of the respective states~2 and~1.  Here, $F_{\rm DM}(q) $ and
$f_{\rm ion}(q)$ are dimensionless form factors following the notation
of~Refs.~\cite{Essig:2011nj,Essig:2012yx}.
Our kinematic setup implies contact interactions with the electron,
$m_V \gg q$, and the free DM-electron scattering cross section is
given by,
\begin{align}
  \bar \sigma_e & = a \frac{16 \pi\,  \alpha\, \alpha_V\, \kappa^2\, \mu_{\chi e}^2 
   }{m_V^4} \simeq 10^{-44} \, \cm^2 \, a 
  \frac{\alpha_V}{\alpha} \left( \frac{\kappa}{10^{-10}} \right)^2 
 \left(  \frac{m}{100\,\keV}\right)^2  \left(\frac{300 \,\keV}{m_V}  \right)^4 \,,  
\end{align}
where $a = 1\, (1/2)$ for fermions (scalars), and assuming
$m\lesssim m_e$ in the second equality.

For liquid scintillator detectors, the process is to a good
approximation described by ionization from an isolated atom from a
shell with principal and angular quantum number $n$ and $l$. The
atomic form factor is then readily evaluated as the Fourier transform
$\chi_{nl}$ of the respective radial bound state wave function,
\begin{align}
\frac{ d \sigma_{nl}\, v }{ d\ln E_{R,e}  }   &  
 = \bar \sigma_e  \frac{(2l+1)\,  m\,
 {p_e'}^2  }{4\, (2\pi)^3\,  \mu_{\chi e}^2  }  \int_{|p_e'-\sqrt{2 m\, ( \Delta m - E_B - E_{R,e} )} |}^{p_e'+ \sqrt{2 m\, ( \Delta m - E_B - E_{R,e} )} } dp' \, p' |\chi_{nl}(p')|^2  \,. 
\end{align} 
Ensuing exemplary recoil spectra are shown in the left panel of
Fig.~\ref{fig:dd}.  The right panel shows the parameter space of
$\kappa$ vs.~$m$ with current sensitivity in direct detection using
the same choice of parameters as in Fig.~\ref{fig:kmbounds} but
equipartitioning the abundance of state 2, $R_0 = 1$. The statistical
procedure for setting the limits follows the one outlined in
Ref.~\cite{Kouvaris:2016afs}; see also Refs.~\cite{Essig:2012yx,An:2014twa,Bloch:2016sjj}.

\bibliographystyle{JHEP}
\tiny
\bibliography{biblio}

\end{document}